# A Closed Form Solution for Reissner's Planar Finite-Strain Beam Using Jacobi Elliptical Functions

Milan Batista

University of Ljubljana, Faculty of Maritime Studies and Transport, Slovenia

milan.batista@fpp.uni-lj.si

(Jun-July-Aug 2015)

## Abstract

In the article we introduce an analytical solution for Reissner's large-deflection finite-strain planar beam subject to an end force and a bending moment. The solution is given in terms of Jacobi elliptical functions. The obtained analytical solution is enhanced with numerical examples. A buckling and post buckling behavior of a beam under axial compressive load applied at the end and subject to various boundary conditions is also discussed in some details. In particular, the buckling factor is derived for each case of the boundary conditions.

## 1 Introduction

In this paper we are concerned with analytical solution of the equations for large deflection of initially straight, weightless, uniformly isotropic, linear elastic beam subject to end load which was in 1972 purposed by Reissner (Reissner, 1972). The theory behind these equations enhanced the well-known Euler-Bernoulli large-deflections beam theory with starching and shearing strain. For convenience we will call a beam which fit this extend theory the Reissner's beam, although the equations are formally identical to the equations, which can be obtained from the planar Cosserat beams theory by assuming linear constitutive relations (Antman, 2005).

Available literature on this very specific subject is relatively rare. The elliptic integrals solution for simple supported extensional beam under axial compressive force has been given by Pflüger (Pflüger, 1950), Stoker (Stoker, 1968) and Magnusson with coauthors (Magnusson et al., 2001). The elliptic integral solution for the Reissner's beam the solution was provided by Humer (Humer, 2013) . It is of order to mention also two other solutions. Goto and coworkers (Goto et al., 1990) published a closed-form solutions for elastic beam with axial and shear deformations, using elliptic integrals. However underlying theory the authors adopt was the Timoshenko beam theory of finite displacements with finite strains and that with small strains. Unlike Hummer solution which involve only elliptic integrals of first and second kind their solution also include elliptic integral of third kind. The elliptic integral solution for extensional beam was also given by Stemple (Stemple, 1990), however, the equations that he integrates were derived from his own beam theory.

This brief review shows that available analytical solutions for extensional and shear-deformable beams are given only in terms of elliptic integrals. A shortcoming of an elliptic integral solution is that it is implicit, meaning that in the formulas for the beam coordinates the independent variable is beam cross-section inclination and not the beam arc length.

The aim of this paper is to provide a solution for the Reissner's beam in terms of Jacobi elliptical functions. For the Euler-Bernoulli beam such solution was proved more suitable for both numerical computation as well as for the analytical treatment than the solution using elliptic integrals (Batista, 2014, 2015a, b; Goss, 2003; Levyakov, 2001; Love, 1944). We will also give an applications of the solution primarily as an indication of its ability.

Before proceeded we note that since we consider only the integration of Reissner's equation we omit reviewing some important topics. Thus, for the history of the large deflections of beams beside mentioned works, we refer to Antman's article (Antman, 1972) , Gorski survey paper (Gorski, 1976) and Goss dissertation (Goss, 2003). For qualitative treatment of the solutions for nonlinear elastic beams the primary reference is Antman's book (Antman, 2005), and for numerical treatment we refer to Saje (Saje, 1991) and Batista and Kosel (Batista and Kosel, 2005) .

## 2 Basic equations

### 2.1 Problem statement

We consider an initially straight Reissner's beam of length $\ell$ , which will be used as unit of length, with one end fixed and a force and a bending moment acting at the other end. In the Cartesian coordinate system $OXY$ the shape of the deformed base curve of the beam is described by the following differential equations (Reissner, 1972), (Eq 14a,14b, 10)

$$\frac{dX}{ds}=(1+\varepsilon)\cos\phi-\gamma\sin\phi, \qquad \frac{dY}{ds}=(1+\varepsilon)\sin\phi+\gamma\cos\phi \tag{1}$$

$$\frac{d\phi}{ds}=\kappa \tag{2}$$

In these equations $X$, Y are coordinates of deformed beam base curve, $\phi$ is angle between $X$ axis and the outward normal to sheared cross section of the beam, $\varepsilon$, $\gamma$, and $\kappa$ are successively axial, transverse and bending strains and parameter $0\le s\le 1$ is length parameter of undeformed beam, measured from the beam immovable end to the beam movable end (Figure 1). Thru this paper we will assume that at initial state beam is on $X$ axis and that the immovable beam end is at the coordinate's origin

$$X(0)=Y(0)=0 \tag{3}$$

Since the beam under load cannot became a point the physical limitation for $\varepsilon$ is (Antman, 2005)

$$1+\varepsilon>0 \tag{4}$$

This condition also prevent that the normal to the sheared cross-section become orthogonal to the base curve.

Beam equilibrium equations are (Reissner, 1972) (Eqa 2*a,b and 3*)

$$\frac{dN}{ds}-\kappa Q=0\,,\qquad \frac{dQ}{ds}+\kappa N=0 \tag{5}$$

$$\frac{dM}{ds}+(1+\varepsilon)Q-\gamma N=0 \tag{6}$$

where $N$ and $Q$ are respectively normal and shear forces with respect to deformed cross section and $M$ is a bending moment.

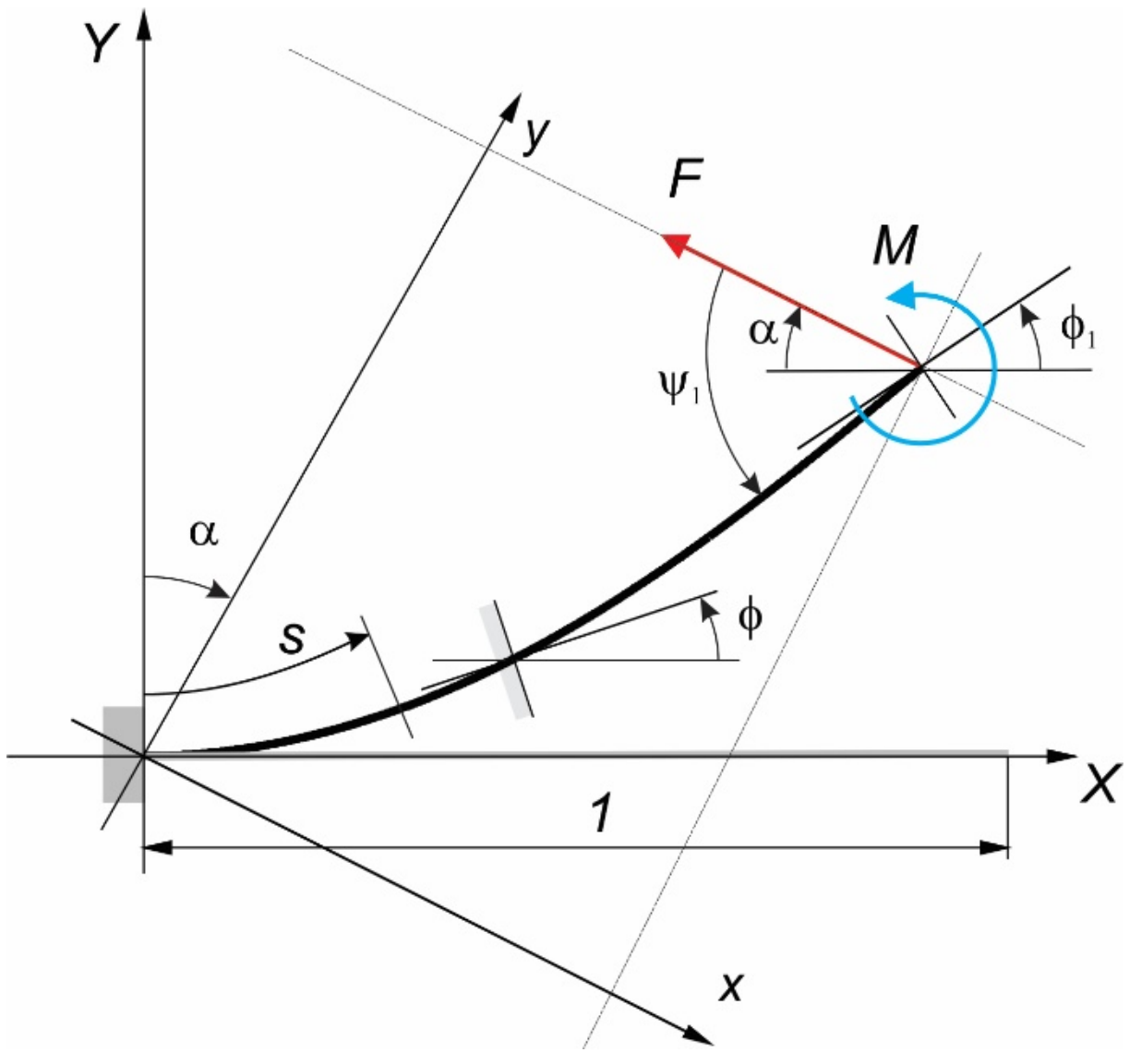

**Figure 1.** Geometry and load of the beam

We assume that the forces and moment are related to deformations by the following linear constitutive equations (Reissner, 1972) (Eqs 32a,32b, 32c)

$$N=EA\varepsilon\,,\qquad Q=GA_s\gamma\,,\qquad M=EI\kappa \tag{7}$$

where $EA$, $GA_s$ and $EI$ are positive constants which represent respectively axial, shear and bending stiffness of the beam. For the interpretation of these equations and future references we refer to Irschik and Gerstmayr (Irschik and Gerstmayr, 2009) and Humer (Humer, 2013)

If the force $F$ is acting on the movable end of beam under clockwise angle $\alpha$ measured from the negative $X$ axis, then the solution of the force equilibrium equations (5) are (Reissner, 1972) (Eq 17 a, b)

$$N=-F\cos(\phi+\alpha),\qquad Q=F\sin(\phi+\alpha) \tag{8}$$

Substituting these expressions for $N$ and $Q$ in constitutive equations $(7)_{1,2}$ we get

$$\varepsilon=-\frac{F}{EA}\cos(\phi+\alpha),\qquad \gamma=\frac{F}{GA_s}\sin(\phi+\alpha) \tag{9}$$

Substituting for $M$ from constitutive relation $(7)_3$ into moment equilibrium equation (6) we obtain, using expression for forces (8),

$$EI\frac{d\kappa}{ds}=-F\left[\left(1+\varepsilon\right)\sin\left(\phi+\alpha\right)+\gamma\cos\left(\phi+\alpha\right)\right] \tag{10}$$

In this way the problem is reduced to the integration of the system of two nonlinear differential equations (2) and (10) for unknowns $\phi$ and $\kappa$. Once these are known $X$ and $Y$ can be obtained by integration of Eqs (1).

### 2.2 Equations transformation

We introduce the new variable $\psi$ defined by

$$\psi\equiv\phi+\alpha \tag{11}$$

which is a counter clockwise angle between the direction of force and inward normal to the beam deformed cross section. Then we rewrite Eqs (9), (2) and (10) in the form

$$\varepsilon=-\frac{\left(1+\nu\right)}{2\lambda^2}\omega^2\cos\psi\,,\qquad \gamma=\frac{\left(1-\nu\right)}{2\lambda^2}\omega^2\sin\psi \tag{12}$$

$$\frac{d\psi}{ds}=\kappa \tag{13}$$

$$\frac{d\kappa}{ds}=-\omega^2\sin\psi\left(1-\frac{\nu\omega^2}{\lambda^2}\cos\psi\right) \tag{14}$$

The non-dimensional parameters in these equation are the load parameter ω, the generalized slenderness ratio $\lambda$ and the stiffness ratio $\nu$ which are defined by (Batista and Kosel, 2005)

$$\omega^2\equiv\frac{F\ell^2}{EI},\qquad \frac{1}{\lambda^2}\equiv\frac{EI}{\ell^2}\left(\frac{1}{GA_s}+\frac{1}{EA}\right)>0\,,\qquad \nu\equiv\frac{GA_s-EA}{GA_s+EA}\in\left[-1,1\right] \tag{15}$$

The parameter $\omega^2$ represent dimensional force while $\omega$ has no physical meaning. In limit when $EA/GA_s=0$ we have $\nu=1$ and $\lambda=\ell\sqrt{EA/EI}$. Such beam is shear-less $\left(\gamma\equiv0\right)$ and allow only stretching. When in limit we have $GA_s/EA=0$ then $\nu=-1$ and $\lambda=\ell\sqrt{GA_s/EI}$ so stretch-less beam $\varepsilon\equiv0$ and only shear deformations are possible. The case $\nu=0$ does not nullify $\varepsilon$ and $\gamma$, however, Eqs (13) and (14) become equations for the Euler-Bernulli beam. Deformations vanish from Eqs (1) when $1/\lambda^2=0$ in limit. In what follows we will use $\nu/\lambda^2=0$ when $\nu=0$ or when $1/\lambda^2=0$.

We note that the ordinary slenderness ratio $\lambda'$ (Humer, 2013; Timoshenko, 1961) and the stiffness ratio $\nu'$ (Humer, 2013) are define by

$$\lambda'\equiv\ell\sqrt{\frac{EA}{EI}}\,,\qquad \nu'\equiv\frac{GA_s}{EA} \tag{16}$$

The connections of these parameters to the present parameters $\lambda$ and $\nu$ are the following

$$\lambda=\frac{\lambda'}{\sqrt{1+1/\nu'}}, \qquad \nu=\frac{\nu'-1}{\nu'+1} \tag{17}$$

For $\nu'=0$ we from these relations obtain $\lambda=0$ and $\nu=-1$. However, in present model $\lambda>0$, because the case $\nu=-1$ is covered by $GA_s/EA=0$ and not by $GA_s=0$. (The later by (7)₂ imply $Q\to 0$ which is incompatible with beam equilibrium equations.) For $\nu'=\infty$ we have $\lambda=\lambda'$ and $\nu=1$. Thus with present parameters we obtain solution for the shear-less beam directly and not thru a limit process.

Using $\varepsilon$ from Eq (12) in condition (4) and taking $\psi=0$ (maximal $\varepsilon$) we obtain upper limit for the load parameter

$$\frac{\omega^2}{\lambda^2}<\frac{2}{1+\nu} \tag{18}$$

In particular for shear-less beam load parameter is limited by the slenderness to $\omega^2<\lambda^2$, while for stretch-less beam the load parameter is unlimited $\omega^2<\infty$.

We next transform Eqs (1) for *X* and *Y* in the following way. By rotating the coordinate system about the origin by $\alpha$ in the clockwise direction we obtain new coordinate system *Oxy*. The transformations between coordinates of the systems are

$$X=x\cos\alpha+y\sin\alpha, \qquad Y=-x\sin\alpha+y\cos\alpha \tag{19}$$

Note that conditions (3) imply

$$x(0)=y(0)=0 \tag{20}$$

Substituting above expressions for *X* and *Y* in to the Eq (1) and using Eq (11) for $\psi$ and Eq (12) for $\varepsilon$ and $\gamma$, we obtain

$$\frac{dx}{ds}=\cos\psi-\frac{\omega^2}{2\lambda^2}\left(1+\nu\cos 2\psi\right), \quad \frac{dy}{ds}=\sin\psi\left(1-\frac{\nu\omega^2}{\lambda^2}\cos\psi\right)=-\frac{1}{\omega^2}\frac{d\kappa}{ds} \tag{21}$$

Now, once we solve differential equations (13) and (14) for unknowns $\psi$ and $\kappa$ we obtain $\phi$ from Eq (11)

$$\phi=\psi-\alpha \tag{22}$$

Further, from Eqs (21) , by integration, using conditions (20), we obtain coordinates

$$x=-\frac{\omega^2}{2\lambda^2}s+\int_0^s\cos\psi\,ds-\frac{\nu\omega^2}{2\lambda^2}\int_0^s\cos 2\psi\,ds \tag{23}$$

$$y=\frac{\kappa_0-\kappa}{\omega^2} \tag{24}$$

where $\kappa_0=\kappa(0)$ . Knowing *x* and *y* we obtain *X* and *Y* from Eqs (19).

### 3 General solution

We eliminate *s* from Eq (14) by putting $\frac{d\kappa}{ds}=\frac{d\kappa}{d\psi}\frac{d\psi}{ds}$ and using Eq (13) for $\frac{d\psi}{ds}$. In this way we obtain

$$\kappa\frac{d\kappa}{d\psi}=-\omega^2\sin\psi\left(1-\frac{\nu\omega^2}{\lambda^2}\cos\psi\right) \tag{25}$$

The result of integration of this equation with respect to $\psi$ and subject to condition $\kappa(\psi_1)=\kappa_1=M\ell/EI$, where $\psi_1\equiv\psi(1)$, is

$$\kappa=\sqrt{2\omega^2\left(\cos\psi-\cos\psi_1\right)\left[1-\frac{\nu\omega^2}{2\lambda^2}\left(\cos\psi_1+\cos\psi\right)\right]+\kappa_1^2} \tag{26}$$

With trigonometric identity

$$\cos\psi=1-2\sin^2\frac{\psi}{2} \tag{27}$$

we can rewrite Eq (26) as

$$\kappa=2\omega\sqrt{\left(\sin^2\frac{\psi_1}{2}-\sin^2\frac{\psi}{2}\right)\left[1-\frac{\nu\omega^2}{\lambda^2}+\frac{\nu\omega^2}{\lambda^2}\left(\sin^2\frac{\psi_1}{2}+\sin^2\frac{\psi}{2}\right)\right]+\left(\frac{\kappa_1}{2\omega}\right)^2} \tag{28}$$

By introducing new variable $\theta$ and two real parameters *k* and *A* given through the relations

$$\sin\frac{\psi}{2}=k\sin\theta\,,\qquad \sin^2\frac{\psi_1}{2}=k^2-A \tag{29}$$

we simplifies Eq (28) to the form

$$\kappa=2\omega k\cos\theta\sqrt{1-\frac{\nu\omega^2}{\lambda^2}+\frac{\nu\omega^2}{\lambda^2}k^2\left(1+\sin^2\theta\right)} \tag{30}$$

where *A* is chosen to satisfy the equation

$$\frac{\nu\omega^2}{\lambda^2}A^2-\left[1+\frac{\nu\omega^2}{\lambda^2}\left(2k^2-1\right)\right]A+\left(\frac{\kappa_1}{2\omega}\right)^2=0 \tag{31}$$

Future, using Eq $(29)_1$, we can express $\kappa$ as

$$\kappa=\frac{d\psi}{ds}=\frac{d\psi}{d\theta}\frac{d\theta}{ds}=\frac{2k\cos\theta}{\sqrt{1-k^2\sin^2\theta}}\frac{d\theta}{ds} \tag{32}$$

so Eq (30) become

$$\frac{d\theta}{ds}=\frac{\tilde{\omega}}{\sqrt{1+m^2}}\sqrt{\left(1-k^2\sin^2\theta\right)\left(1+m^2\sin^2\theta\right)} \tag{33}$$

Here the parameters $m^2$ and $\omega_1$ are defined by

$$m^2 \equiv \frac{\frac{\nu\omega^2}{\lambda^2}k^2}{1-\frac{\nu\omega^2}{\lambda^2}\left(1-k^2\right)}, \qquad \tilde{\omega} \equiv \omega\sqrt{1+\frac{\nu\omega^2}{\lambda^2}\left(2k^2-1\right)} \tag{34}$$

With variable $u$ defined by

$$u \equiv \sin\theta \tag{35}$$

we rewrite Eq (33) in the algebraic form

$$\frac{du}{ds} = \frac{\tilde{\omega}}{\sqrt{1+m^2}}\sqrt{\left(1-u^2\right)\left(1-k^2u^2\right)\left(1+m^2u^2\right)} \tag{36}$$

Further simplification of this equation is obtained by introducing yet another variable $t$ defined by Gauss type transformation (Groebner and Hofreiter, 1961)

$$u \equiv \frac{t}{\sqrt{1+m^2\left(1-t^2\right)}} \tag{37}$$

Substituting this in Eq (36) , we get

$$\frac{dt}{ds} = \tilde{\omega}\sqrt{\left(1-t^2\right)\left(1-\tilde{k}^2t^2\right)} \tag{38}$$

where $\tilde{k}$ is a constant defined by

$$\tilde{k}^2 \equiv \frac{k^2+m^2}{1+m^2} = k^2\frac{1+\frac{\nu\omega^2}{\lambda^2}k^2}{1+\frac{\nu\omega^2}{\lambda^2}\left(2k^2-1\right)} \tag{39}$$

As is well-known the general solution of Eq (38) is

$$t\left(s\right) = \mathrm{sn}\left(\tilde{\omega}s+C,\tilde{k}\right) \tag{40}$$

were *sn* is Jacobi elliptic function and *C* is constant of integration. From this we obtain by consecutive substitutions into Eq (37) , then obtained expression into Eq (35) , and finally into Eq (29), the final solution of Eqs (13) and (14)

$$\psi = 2\arcsin\left[k\frac{\mathrm{sn}\left(\tilde{\omega}s+C,\tilde{k}\right)}{\sqrt{1+m^2\mathrm{cn}^2\left(\tilde{\omega}s+C,\tilde{k}\right)}}\right] \tag{41}$$

$$\kappa = \frac{2\tilde{\omega}k\sqrt{1+m^2}\,\mathrm{cn}\left(\tilde{\omega}s+C,\tilde{k}\right)}{1+m^2\mathrm{cn}^2\left(\tilde{\omega}s+C,\tilde{k}\right)} \tag{42}$$

where *cn* is Jacobi elliptic function. Using these expressions for $\psi$ and $\kappa$ we from Eqs (23) and (24) obtain coordinates. Both integrals in Eq (23) was found by the Maple program. We note that complete elliptic integral of third kind vanish from expression for *x* once obtained result is simplified. The final result can be written in the form

$$x=-\frac{(1-\nu)\omega^2}{2\lambda^2}s+\frac{2\tilde{\omega}}{\omega^2}\left\{\left(\frac{E(\tilde{k})}{K(\tilde{k})}-\frac{1}{2}\right)\tilde{\omega}s+Z(\tilde{\omega}s+C,\tilde{k})-Z(C,\tilde{k})\right.$$
$$\left.-m^2\left[\frac{\mathrm{sn}(\tilde{\omega}s+C,\tilde{k})\mathrm{cn}(\tilde{\omega}s+C,\tilde{k})\mathrm{dn}(\tilde{\omega}s+C,\tilde{k})}{1+m^2\mathrm{cn}^2(\tilde{\omega}s+C,\tilde{k})}-\frac{\mathrm{sn}(C,\tilde{k})\mathrm{cn}(C,\tilde{k})\mathrm{dn}(C,\tilde{k})}{1+m^2\mathrm{cn}^2(C,\tilde{k})}\right]\right\} \tag{43}$$

$$y=\frac{2k\tilde{\omega}\sqrt{1+m^2}}{\omega^2}\left[\frac{\mathrm{cn}(C,\tilde{k})}{1+m^2\mathrm{cn}^2(C,\tilde{k})}-\frac{\mathrm{cn}(\tilde{\omega}s+C,\tilde{k})}{1+m^2\mathrm{cn}^2(\tilde{\omega}s+C,\tilde{k})}\right]$$

Here *K* and *E* are complete elliptic integrals of first and second kind, respectively, and *dn* and *Z* are Jacobi's elliptic functions.

The remaining unknown of some interest is the length of deformed beam *L* which is given by

$$L=\int_0^1\sqrt{\left(\frac{dX}{ds}\right)^2+\left(\frac{dY}{ds}\right)^2}\,ds=\int_0^1\sqrt{(1+\varepsilon)^2+\gamma^2}\,ds \tag{44}$$

In general this integral has no closed form solution. However, there is a special case $\gamma\equiv 0$ when it become

$$L=\int_0^1(1+\varepsilon)ds \tag{45}$$

Integration yield the following expression, valid for $\nu=1$,

$$L=1-\left(1+\frac{2k^2}{m^2}\right)\frac{\omega^2}{\lambda^2}+\frac{2k^2}{m^2\tilde{\omega}}\frac{\omega^2}{\lambda^2}\left[\Lambda\left(\tilde{\omega}+C,\frac{m^2}{1+m^2},\tilde{k}\right)-\Lambda\left(C,\frac{m^2}{1+m^2},\tilde{k}\right)\right] \tag{46}$$

where $\Lambda(z,\alpha,k)\equiv\Pi(\mathrm{am}z,\alpha,k)$ , $\Pi$ is elliptic integral of third kind (Lawden, 1989) and *am* is Jacobi's amplitude function (Reinhardt and Walker, 2010).

## 4 Properties of the solution

### 4.1 Solution domain

The solution given by Eqs (41), (42), (43) contains three parameters $\lambda$ , $\nu$ , $\omega$ and two constants *C* and $k$ by means of which we can calculate parameters $m^2$ , $\tilde{\omega}$ , $\tilde{k}$ by using Eqs (34) and (39) respectively. Since we interested only in real solution we assume that *k* is real and that

$$1+\frac{\nu\omega^2}{\lambda^2}\left(2k^2-1\right)>0 \tag{47}$$

This inequality ensures that $\tilde{\omega}$ is real and also that $1+m^2>0$, both of which are easily verified by Eq (34). From inequality (18) it follows that $\frac{\nu\omega^2}{\lambda^2}<1$. Using this and Eq (47) we can derive the bounds on $m^2$, $\tilde{\omega}$, $\tilde{k}$

$$-1<m^2<1, \quad 0<\tilde{\omega}<\omega\begin{cases}\sqrt{2\left(1-k^2\right)} & k^2<1\\ k\sqrt{2} & k>1\end{cases}, \quad -\infty<\tilde{k}^2<\begin{cases}\frac{1}{2}\left(1+k^2\right) & k^2<1\\ \dfrac{k^4}{2k^2-1} & k^2>1\end{cases} \tag{48}$$

The inequalities (18), (47) and contour lines of $m^2$ and $\tilde{k}$, are shown graphically in Fig 2. On region $\frac{\nu\omega^2}{\lambda^2}\geq 1$ the solution does not exist due to inequality (18). On the region $0\leq\frac{\nu\omega^2}{\lambda^2}<1$ $k$ is unlimited, while for $\frac{\nu\omega^2}{\lambda^2}<0$ the upper limit for $k$ is given by boundary of inequality (47). Note that $\tilde{k}$ become pure complex number when $\frac{\nu\omega^2}{\lambda^2}<-1$ and $k^2\frac{\nu\omega^2}{\lambda^2}+1<0$.

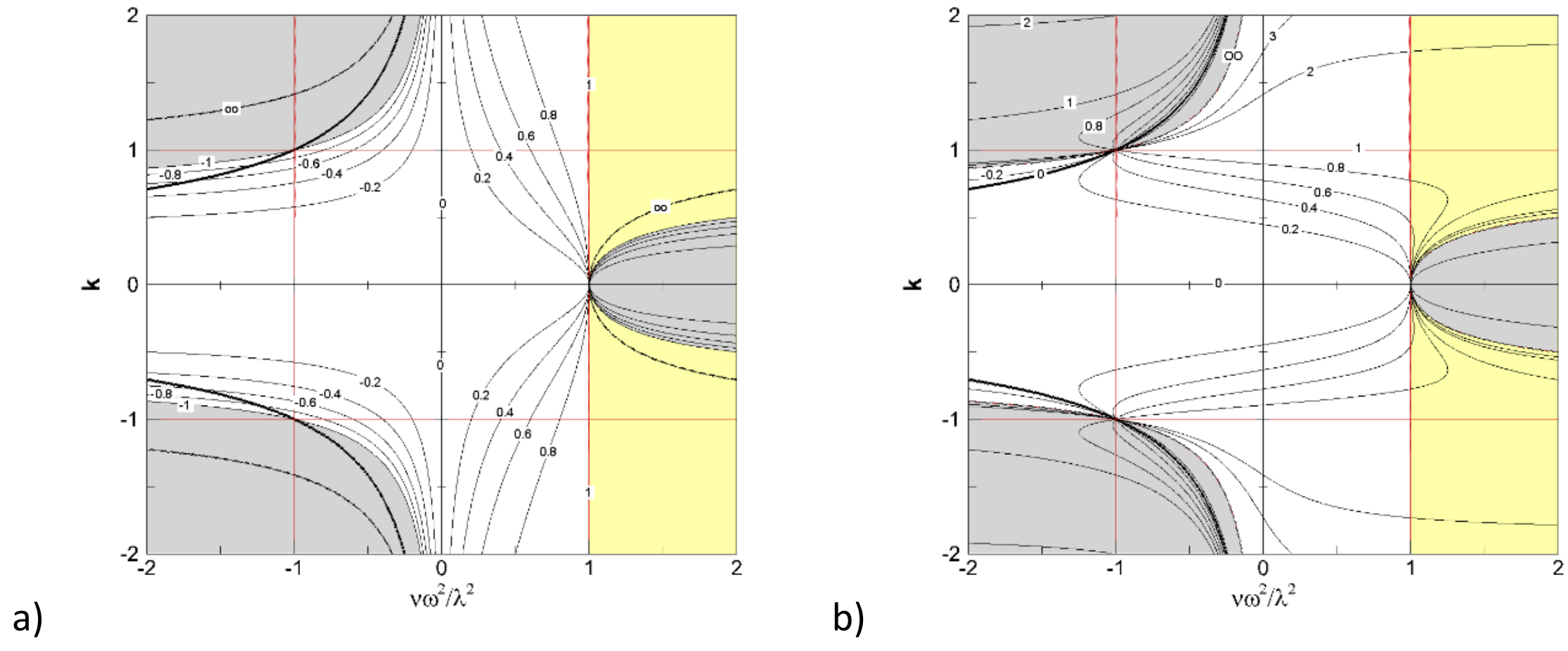

**Figure 2.** The graph of inequality (47) and contour lines of $m^2$ (a) and $\tilde{k}^2$ (b). On dark (gray) shaded areas the inequality fails. On light (yellow) shaded area the inequality (47) is overriding by inequality (18). On the bold solid line $\tilde{k}=0$ .

The formulas for $\psi$, $\kappa$, $x$ and $y$ for the case when $\tilde{k}\geq 1$ or when $\tilde{k}$ is pure complex number can be easily derive from Eqs (41), (42), (43) with the help of the formulas which relate Elliptic integrals and Jacobi elliptic functions with moduli outside interval $[0,1)$ or purely imaginary moduli to one with moduli in this interval. These formulas are listed in Appendix. Such explicate formulas can be useful in an analytical investigations, however, for numerical calculations one can only implements computations of elliptic functions for different range of moduli. From the Eqs (42),(106),(112) for $\kappa$ we can like in the case of elastica (Batista, 2014; Goss, 2003; Love, 1944) classify three different types of underlying curves: for $\tilde{k}<1$ the underlying curve is inflectional (Fig. 4a), for $\tilde{k}=1$ the curve is homoclinic (Fig. 4b) and for $\tilde{k}>1$ the curve is non-inflectional (Fig. 4c)

### 4.2 Formulas for *k*

The solution do not depend on $\kappa_1$, which appears in Eq (31). To see that we substitute *A* from Eq (29) 2 into Eq (31). In this way we obtain

$$\frac{\nu\omega^2}{\lambda^2}k^4+\left(1-\frac{\nu\omega^2}{\lambda^2}\right)k^2-\sin^2\frac{\psi_1}{2}\left(1-\frac{\nu\omega^2}{\lambda^2}\cos^2\frac{\psi_1}{2}\right)-\left(\frac{\kappa_1}{2\omega}\right)^2=0 \tag{49}$$

This relation is automatically satisfied by $\psi_1=\psi(1)$ from Eq (41) and $\kappa_1=\kappa(1)$ from Eq (42). In fact, it is the identity for any pair $(\psi,\kappa)$. However, we can use the relation to calculate *k* when $\psi_1$ and $\kappa_1$ are given. Solving Eq (49) for $k^2$ and retaining only the solution for which $k^2$ is bounded when $\frac{\nu}{\lambda^2}=0$, we obtain

$$k^2=\frac{2\left[\sin^2\frac{\psi_1}{2}\left(1-\frac{\nu\omega^2}{\lambda^2}\cos^2\frac{\psi_1}{2}\right)+\left(\frac{\kappa_1}{2\omega}\right)^2\right]}{1-\frac{\nu\omega^2}{\lambda^2}+\sqrt{\left(1-\frac{\nu\omega^2}{\lambda^2}\cos\psi_1\right)^2+\frac{\nu\kappa_1^2}{\lambda^2}}} \tag{50}$$

When $\frac{\nu}{\lambda^2}=0$ the above expression reduce to

$$k^2=\sin^2\frac{\psi_1}{2}+\left(\frac{\kappa_1}{2\omega}\right)^2 \tag{51}$$

which is relation for elastica (Batista, 2014). Alternatively, we can write Eq (49) in the factorized form

$$\left(\frac{\kappa_1}{2\omega}\right)^2=\left(k^2-\sin^2\frac{\psi_1}{2}\right)\left[1+\frac{\nu\omega^2}{\lambda^2}\left(k^2-\cos^2\frac{\psi_1}{2}\right)\right] \tag{52}$$

When $\kappa_1=0$ then above relation is satisfied if

$$k^2=\sin^2\frac{\psi_1}{2} \tag{53}$$

This relation is valid for any inflection point. When $k^2=1$ then $\kappa_1$ and $\omega$ are not independent. As follows from Eq (52), they are related by

$$\kappa_1=\pm2\omega\cos\frac{\psi_1}{2}\sqrt{1+\frac{\nu\omega^2}{\lambda^2}\sin^2\frac{\psi_1}{2}} \tag{54}$$

### 4.3 Symmetry

By replacing $k$ in Eqs (41), (42), (43) by $-k$ we obtain

$$\psi(-k)=-\psi(k),\qquad \kappa(-k)=-\kappa(k),\qquad x(-k)=x(k),\qquad y(-k)=-y(k) \tag{55}$$

This is solution which describe the beam which is symmetric with respect to *x* axis. However in order to obtain a symmetric shape in the system *OXY* we must also replace $\alpha$ in Eq (19) by $-\alpha$ . Then we have, from Eqs (19),

$$X(-k,-\alpha)=X(k,\alpha), \quad Y(-k,-\alpha)=-Y(k,\alpha) \tag{56}$$

### 4.4 Some special cases

When $\frac{1}{\lambda^2}=0$ then we have, from Eqs (34) and (39) ,

$$m^2=0, \quad \tilde{\omega}=\omega, \quad \tilde{k}=k \tag{57}$$

In this case Eqs (41), (42), (43) reduce to known solution for elastica (Batista, 2014)

$$\psi=2\arcsin\left[k\,\mathrm{sn}(\omega s+C,k)\right], \quad \kappa=2\omega k\,\mathrm{cn}(\omega s+C,k) \tag{58}$$

$$x=\frac{2}{\omega}\left[\left(\frac{E(k)}{K(k)}-\frac{1}{2}\right)\omega s+Z(\omega s+C,k)-Z(C,k)\right], \quad y=\frac{2k}{\omega}\left[\mathrm{cn}(C,k)-\mathrm{cn}(\omega s+C,k)\right] \tag{59}$$

In the case when $k=0$ we obtain by means of Eqs (34) and (39)

$$m^2=0, \quad \tilde{\omega}=\omega\sqrt{1-\frac{\nu\omega^2}{\lambda^2}}, \quad \tilde{k}=0 \tag{60}$$

Hence the solution given by Eq (41), (42), (43) reduces to well-known solution for a straight beam under axial force

$$\psi=0, \quad \kappa=0, \quad x=\left[1-\frac{(1+\nu)\omega^2}{2\lambda^2}\right]s, \quad y=0 \tag{61}$$

When $\omega=0$ then we obtain the solution for the pure bending of the beam. From Eqs (34) and (39) we have

$$m^2=0, \quad \tilde{\omega}\approx\omega, \quad \tilde{k}=k \tag{62}$$

It follows from Eq (50) that $k\to\infty$ as $\omega\to 0$ , however, in this limit we have $k\omega=\frac{\kappa_1}{2}$ . With these results and expansion of Z function for small modulus (Batista, 2014) (Eq 57), we find , on using Eq (41), (42), (43)

$$\psi=\kappa_1 s+2C, \quad \kappa=\kappa_1, \quad x=\frac{\sin(\kappa_1 s+2C)-\sin(2C)}{\kappa_1}, \quad y=\frac{\cos(2C)-\cos(\kappa_1 s+2C)}{\kappa_1} \tag{63}$$

This results shows that the beam subject to pure bending deforms into a circular arc. In this case $\alpha$ is indeterminate and represent rigid body rotation.

## 5 Numerical examples

In this section we will use the present solution for calculation of a shape of deformed cantilever under end load. The boundary conditions of the problem are

$$\phi(0)=0 \text{ (clamped end)}, \qquad \kappa(1)=\kappa_1 \text{ (free end)} \tag{64}$$

where $\kappa_1$ is expressed in terms of moment $M$ by $\kappa_1 = M\ell/EI$. We assume that parameters $\lambda$, $\nu$, $\omega$, $\kappa_1$ are given and that the unknowns of the problem are $k$, $C$ and possible $\alpha$. We will consider only the nontrivial cases when $k \neq 0$.

### 5.1 Cantilever under follower force

As the first example we consider a cantilever subject only to a follower force. The direction of the force with respect to beam is in this case $\psi_1 = \psi(1)$, therefore, $k$ is given by Eq (53). By means of Eqs (11) and (41) and the boundary condition $\phi(0)=0$ we find the expression for unknown $\alpha$

$$\alpha = 2\sin^{-1}\left( k\frac{\mathrm{sn}\left(C,\tilde{k}\right)}{\sqrt{1+m^2\mathrm{cn}^2\left(C,\tilde{k}\right)}} \right) \tag{65}$$

Substituting $\kappa$ given by Eq (42) into the boundary condition $\kappa(1)=0$, we obtain the equation for unknown $C$

$$\frac{2\tilde{\omega}k\sqrt{1+m^2}\,\mathrm{cn}\left(\tilde{\omega}+C,\tilde{k}\right)}{1+m^2\mathrm{cn}^2\left(\tilde{\omega}+C,\tilde{k}\right)} = 0 \tag{66}$$

This equations is satisfied if $\mathrm{cn}\left(\tilde{\omega}+C,\tilde{k}\right)=0$, and therefore in particular for

$$C = -\tilde{\omega} + K\left(\tilde{k}\right) \tag{67}$$

In this way we obtain an explicit expressions for all three unknowns' $k$, $\alpha$ and $C$. The results obtained by the present analytical methods presented in Table 1 and in Fig 3 agrees with those obtained by numerical integration (Batista, 2013).

**Table 1.** Comparison of results for cantilever subject to follower force when $\psi_1 = 90^0$, $\lambda = 15$, $\omega = 10$. The analytic values were obtained by the Maple program with $digits = 14$. Numerical results are from Batista (Batista, 2013).

| $\nu$ | C | $\alpha(\deg)$ | $X(1)$ | $Y(1)$ | $\phi(1)(\deg)$ | $\kappa(0)$ | $L$ | method | Shape in Fig 2 |
|---|---|---|---|---|---|---|---|---|---|
| -1 | -8.230109 | -78.868401 | -0.028481 | 0.183947 | 168.868401 | 6.345803 | - | Analytic | a |
| | | | -0.02848 | 0.18395 | 168.8684 | 6.3458 | 1.0626 | Numeric | |
| 0 | -8.145925 | -54.251222 | 0.066038 | 0.276759 | 144.251222 | 10.809555 | - | Analytic | b |
| | | | 0.06604 | 0.27676 | 144.2512 | 10.8096 | 0.90883 | Numeric | |
| 1 | -8.038243 | -13.617960 | 0.283974 | 0.195815 | 103.617960 | 12.344909 | 0.785552 | Analytic | c |
| | | | 0.28397 | 0.19582 | 103.618 | 12.3449 | 0.78555 | Numeric | |

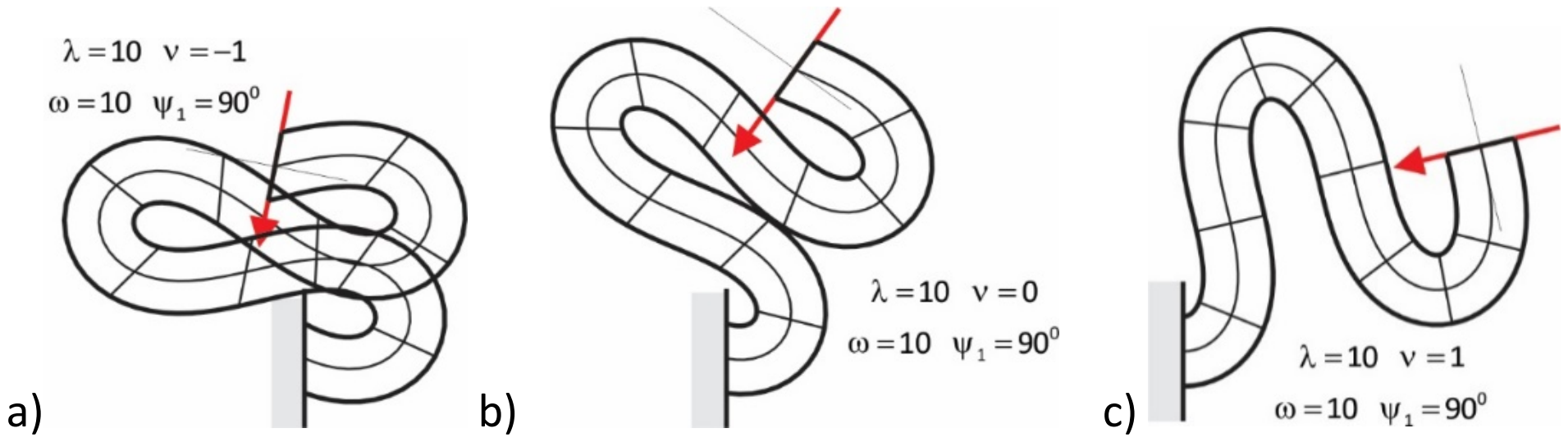

**Figure 3.** Equilibrium shapes of cantilever for the cases in Table 1

### 5.2 Follower force and moment

As second example we consider a cantilever subject to the end follower force and the end moment. We therefore have $\kappa_1 \neq 0$ and thus $k$ is given by Eq (50) while $\alpha$ is given by Eq (65). From the boundary condition $\kappa(1) = \kappa_1$, on using Eq (42), we obtain equation for $C$

$$\frac{2\tilde{\omega} k\sqrt{1+m^2}\,\mathrm{cn}\left(\tilde{\omega}+C,\tilde{k}\right)}{1+m^2\mathrm{cn}^2\left(\tilde{\omega}+C,\tilde{k}\right)} = \kappa_1 \tag{68}$$

From this quadratic equation for $\mathrm{cn}\left(\tilde{\omega}+C,\tilde{k}\right)$ we find

$$C = -\tilde{\omega} + \mathrm{cn}^{-1}\left(\frac{\kappa_1}{\tilde{\omega} k\sqrt{1+m^2} + \sqrt{\tilde{\omega}^2 k^2 - m^2\left(\tilde{\omega}^2 k^2 - \kappa_1^2\right)}}, \tilde{k}\right) \tag{69}$$

Again, we obtain the explicit expressions for all three unknowns' $k$, $\alpha$ and $C$. We note that when $k = 1$ then $\kappa_1$ is given by Eq (54).

For comparison with present analytical solution we conduct numerical integration of Eqs. (13), (14) and (21) for three cases by using the Shvartsman method (Batista, 2013; Shvartsman, 2007). Again, the agreement between results of analytical and numerical solution are excellent (Table 2).

**Table 2**. Comparison between analytical and numerical solutions for cantilever subject to follower force and end moment when $\lambda=10$ , $\nu=-1$ and $\psi_1=\pi/3$ . The analytic solution were obtained by the Maple program with *digits* $=14$ . Numerical solution were obtained by routine *dopri5 (Hairer et al., 1987)* with all tolerances set to $10^{-7}$ . *Diff* is difference between solutions $\times 10^{-8}$ .

| | a) inflectional | | | b) homoclinic | | | c) non-inflectional | | |
|---|---|---|---|---|---|---|---|---|---|
| | Analytic | Numeric | diff | Analytic | Numeric | diff | Analytic | Numeric | diff |
| $\omega$ | 4 | | | 5 | | | 2 | | |
| $\kappa_1$ | 1 | | | 8.3852549 | 8.3852549 | 0.0 | 5 | | |
| $C$ | -2.7445301 | | | -3.8489152 | | | -2.11549168 | | |
| $\alpha/\pi$ | -0.1910131 | -0.1910130 | 7.2 | -0.9765103 | -0.9765101 | 18.9 | -1.11115111 | -1.11115111 | 0.7 |
| $\kappa_0$ | -3.5373453 | -3.5373452 | 5.0 | 0.3195424 | 0.3195399 | -249.7 | 3.68565782 | 3.68565778 | -4.5 |
| $x(1)$ | 0.6340054 | 0.6340056 | 19.6 | -0.4600979 | -0.4600979 | 9.3 | -0.01609271 | -0.01609272 | -1.5 |
| $y(1)$ | -0.2835841 | -0.2835840 | 7.4 | -0.3226285 | -0.3226285 | -1.6 | -0.32858554 | -0.32858553 | 1.4 |
| $\psi_1/\pi$ | 0.5243464 | 0.5243462 | -18.7 | 1.3098436 | 1.3098436 | -2.9 | 1.44448445 | 1.44448447 | 2.4 |
| $X(1)$ | 0.6833802 | 0.6833804 | 14.8 | 0.4826326 | 0.4826326 | 7.9 | -0.09730017 | -0.09730014 | 2.6 |
| $Y(1)$ | 0.1239926 | 0.1239927 | 1.8 | 0.2878282 | 0.2878280 | -26.3 | 0.31426122 | 0.31426121 | -0.6 |

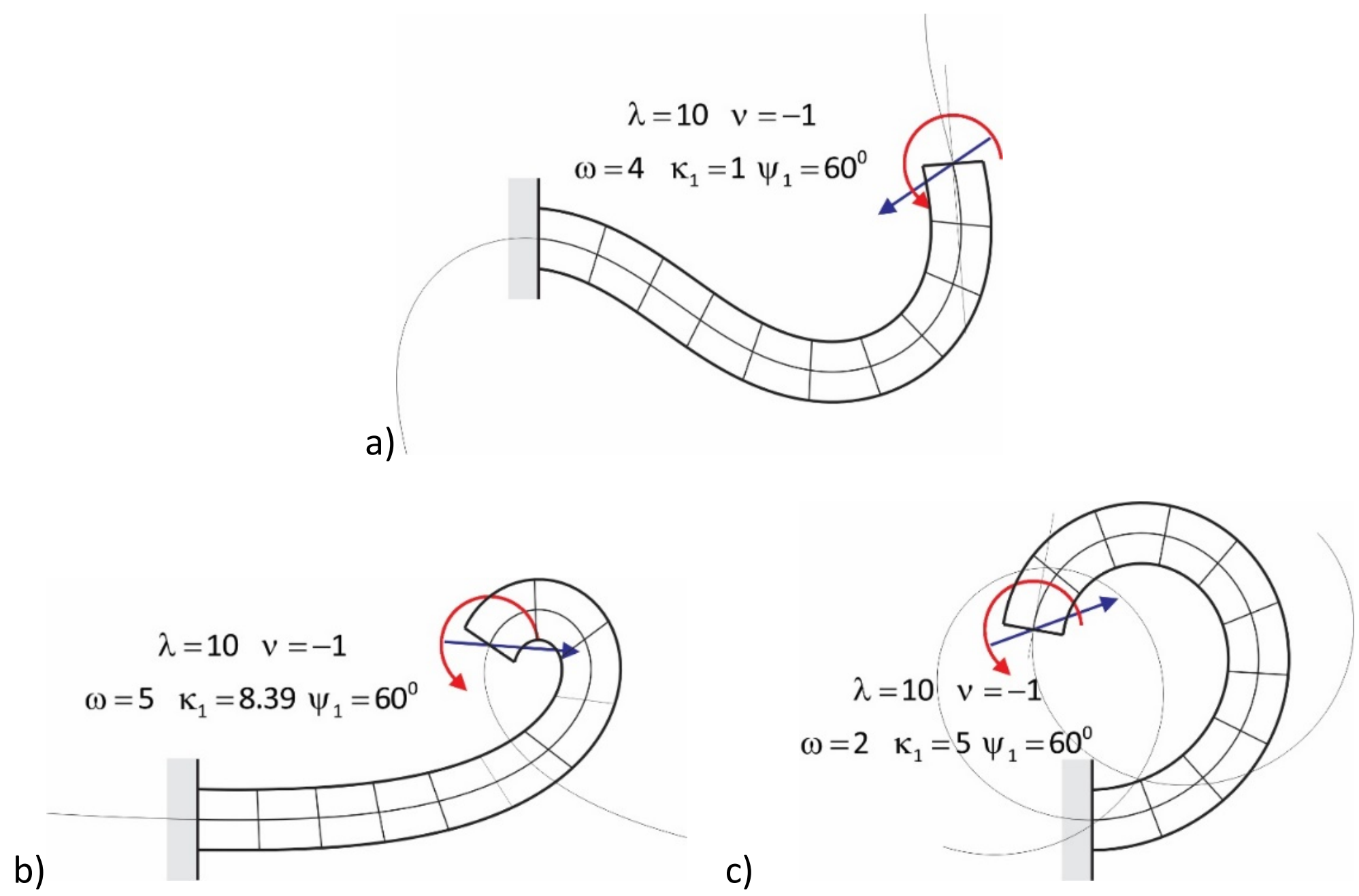

**Figure 4**. Beam equilibrium shapes for the cases in Table 2

### 5.3 Cantilever under constant force

In this case $\alpha$ is given. From the boundary conditions $\phi(0)=0$, on using Eqs (11) and (41) , we obtain unknown *C*

$$C=\operatorname{sn}^{-1}\left(\sin\frac{\alpha}{2}\sqrt{\frac{1+m^2}{k^2+m^2\sin^2\frac{\alpha}{2}}},\tilde{k}\right) \tag{70}$$

By means of Eq (42) the boundary condition $\kappa(1)=0$ yield equation $\operatorname{cn}\left(\tilde{\omega}+C,\tilde{k}\right)=0$ which is identically satisfied by

$$\tilde{\omega}+C=(2n-1)K\left(\tilde{k}\right) \quad (n=1,2,\ldots) \tag{71}$$

This is transcendental equation for unknown $k$ which can be solved by numerical methods. Alternatively, we can in this case take $k=\sin\frac{\psi_1}{2}$ so the equation is to be solved for $\psi_1$.

For numerical calculation we consider the example due to Saje (Saje, 1991) and was also treated by present author (Batista and Kosel, 2005). Agreement between results is perfect (Table 3 and Fig 5).

**Table 3.** Cantilever beam with $\alpha=90^0$, $F=10$, $L=1$, $EA=20^{21}$, $EI=10$ (Saje, 1991). Numerical values are from (Batista and Kosel, 2005) . *Diff* is difference between solutions $\times10^{-8}$

| $GA_s$ | $\lambda$ | $\psi_1/\pi$ | $1-X(1)$ | | | $Y(1)$ | | |
|---|---|---|---|---|---|---|---|---|
| | | | Analytic | Numeric | Diff | Analytic | Numeric | Diff |
| 5x10^20 | 6.9x10^22 | 0.646852887 | 0.056433236 | 0.05643324 | 0.37 | 0.301720774 | 0.30172077 | -0.38 |
| 500 | 7.0711 | 0.645916287 | 0.061315658 | 0.06131566 | 0.16 | 0.317813874 | 0.31781387 | -0.39 |
| 50 | 2.2361 | 0.637850849 | 0.103284917 | 0.10328492 | 0.32 | 0.465413303 | 0.46541330 | -0.35 |
| 10 | 1.0000 | 0.609124015 | 0.252136606 | 0.25213661 | 0.37 | 1.167095878 | 1.16709588 | 0.16 |
| 5* | 0.7071 | 0.585257743 | 0.376121399 | 0.37612140 | 0.09 | 2.104087473 | 2.10408747 | -0.28 |

* $\tilde{k}$ is pure imaginary number

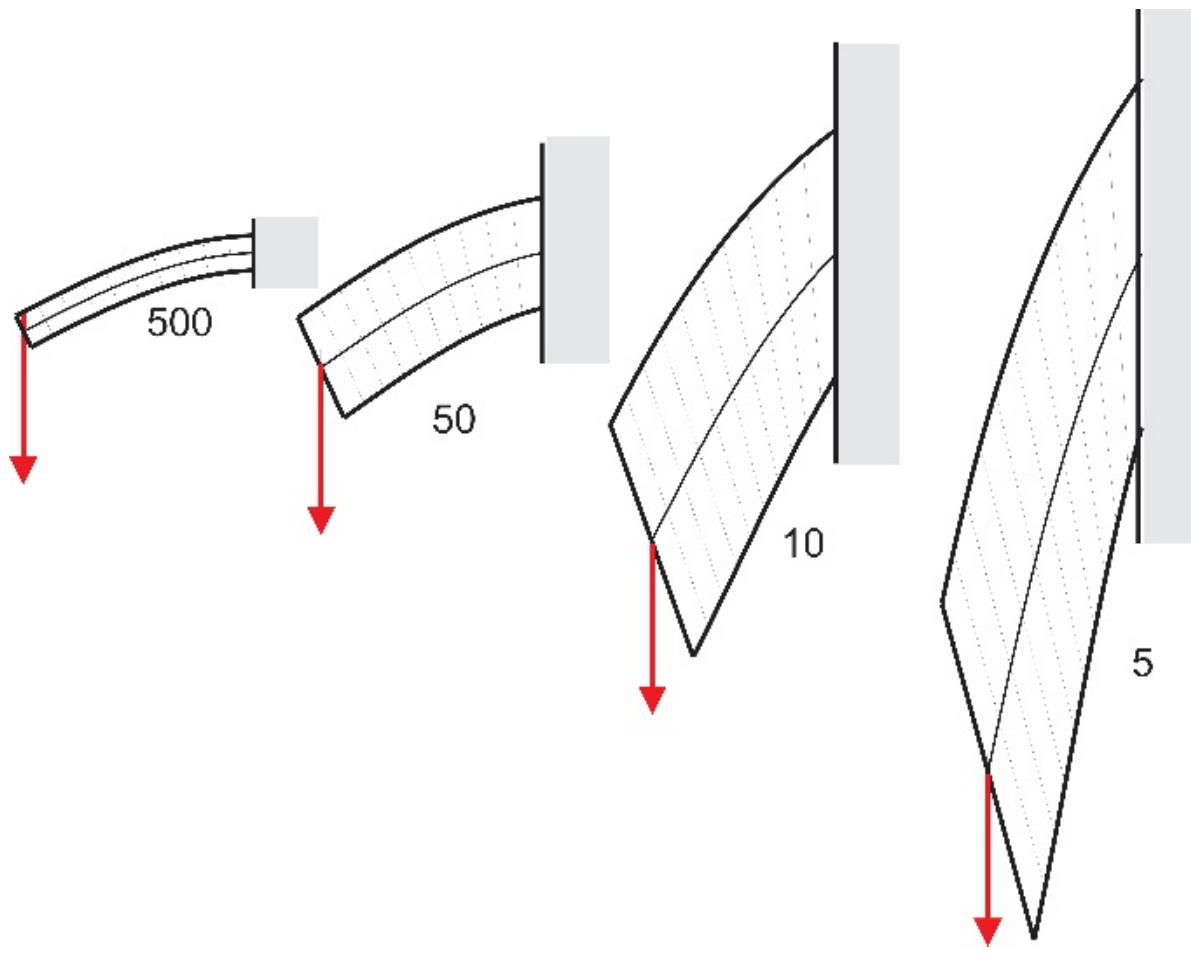

**Figure 5**. Shapes of cantilever for $GA_s\in\{5,10,50,500\}$ for the cases in Table 3. The increasing beam thickness $b\equiv1/\lambda$ is artificial and demonstrate an effect of decreasing of $\lambda$ with decreasing $GA_s$.

## 6 Beam under compressive axial force

In this section we will demonstrate an analytical capabilities of the present solution by treating the classical problem of the beam under axial compressive force subject to various boundary conditions (Figure 6). Our main objective is derivation of critical (buckling) factor $\beta$ defined by

$$\beta \equiv \frac{\omega^2}{\pi^2} \tag{72}$$

With $\beta$ the critical force is expressed in the following way (Ziegler, 1977)

$$F_c = \beta \frac{\pi^2 EI}{\ell^2} \tag{73}$$

We will also consider post buckling behavior of doubly spurted beams in some details. As is well-known, a doubly supported elastica undergoes the secondary loss of stability with increasing force (Levyakov, 2001). We make conjecture that for Reissner's beam secondary loss of stability occurs under the same condition as for elastica. We will call corresponded $\beta$ the secondary critical factor. We will consider only first buckling mode since higher buckling modes are most probably unstable and thus has no practical values.

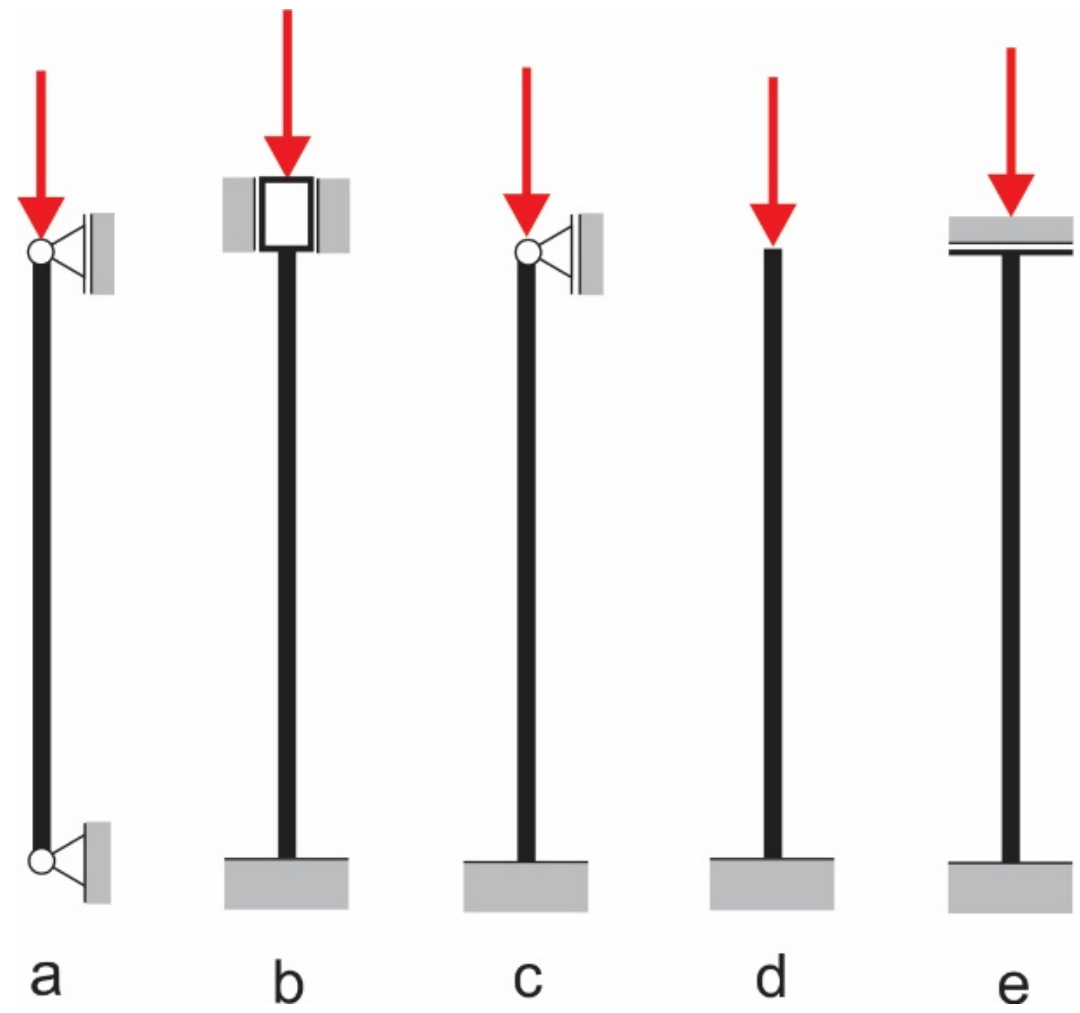

**Figure 6.** Load cases: a) simply supported (SS) beam, b) clamped (CC) beam, c) clamped-hinged (CS) beam, d) cantilever, e) cantilever with guided end

### 6.1 Simply supported beam

The boundary conditions at beam's simpley supported ends are

$$\kappa(0) = \kappa(1) = 0 \;, \quad Y(1) = 0 \tag{74}$$

From $\kappa(0) = 0$ we obtain, on using of Eq (42), $\mathrm{cn}\left(C, \tilde{k}\right) = 0$. This is satisfied by

$$C = K(\tilde{k}) \tag{75}$$

Similarly, from $\kappa(1) = 0$ we get $\mathrm{cn}(\tilde{\omega} + C, \tilde{k}) = 0$ which is satisfied if

$$\tilde{\omega} = 2nK(\tilde{k}) \quad (n = 1, 2, \ldots) \tag{76}$$

This is characteristic equation which relate $k$ and $\omega$ when the beam is in equilibrium. Figure 7 shows an examples of pitchfork bifurcation diagrams of Eq (76) . For $\nu > 0$ the bifurcation point become unstable by lowering the slenderness (Fig 7b). For extensible beam this was observed by Magnuson and coauthors (Magnusson et al., 2001). The transition point can be calculated analytically in the following way. If we denote $f \equiv \tilde{\omega} - 2K(\tilde{k})$ and $F \equiv \frac{\partial^2 f}{\partial k^2}$ then critical slenderness $\lambda_c$ is obtained as solution of this system of equations for $k = 0$. Performing the calculation we obtain $\lambda_c = 4\pi\sqrt{\nu/3}$ . For $\nu = 1$ this is $\lambda_c = 4\pi/\sqrt{3} \approx 2.3096\pi$ which agree with value given by Magnusson and coauthors (Magnusson et al., 2001)

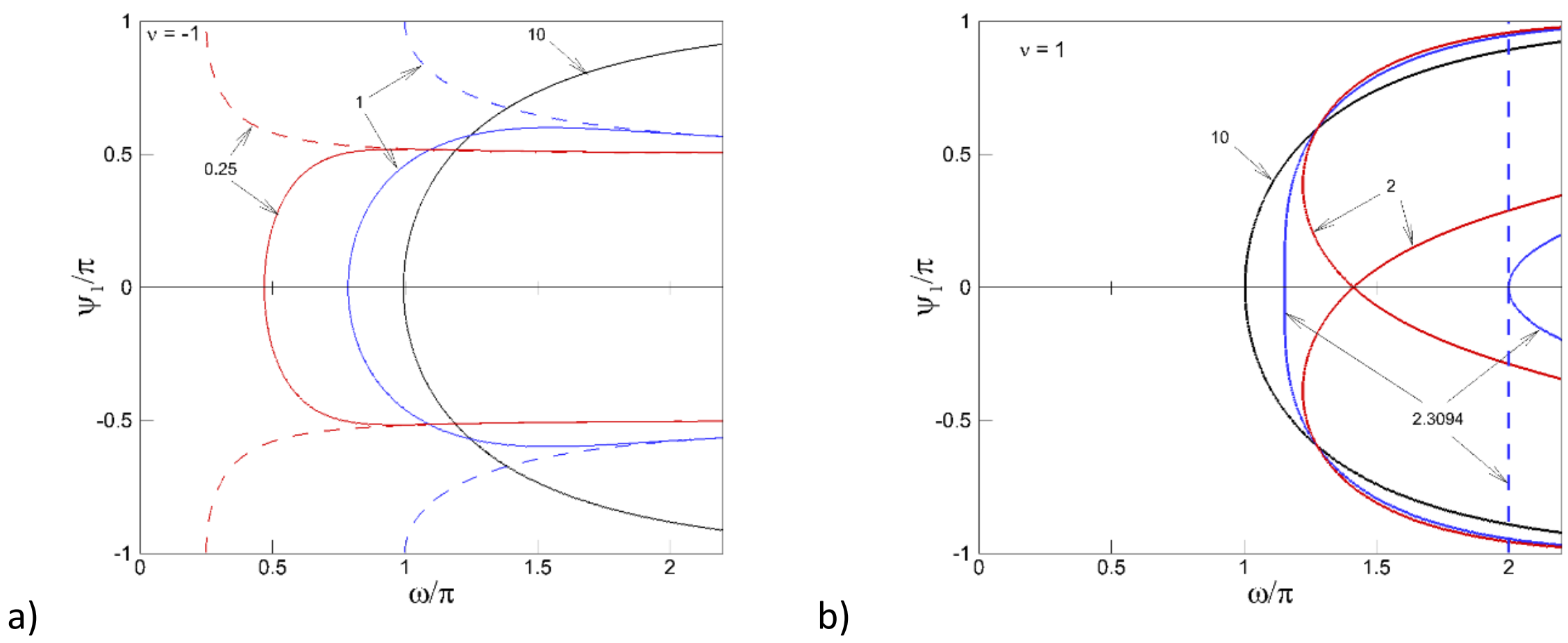

**Figure 7.** Bifurcation diagram of Eq (76) for simply supported beam for $n = 1$ . a) $\nu = -1$ , $\lambda/\pi \in \{0.25, 1, 10\}$ , the dash line is boundary imposed by inequality (47) . b) $\nu = 1$ , $\lambda/\pi \in \{2, 2.3094, 10\}$ , the dashed lines is boundary imposed by inequality (18).

In order to obtain critical force we set $k = 0$ in Eq (76). This yields equation $\omega\sqrt{1 + \frac{\nu\omega^2}{\lambda^2}} = n\pi$ for $\omega$. From the solution of this equation for $n = 1$ we find the following critical factor

$$\beta = \frac{\omega^2}{\pi^2} = \frac{2}{1 + \sqrt{1 - \frac{4\pi^2\nu}{\lambda^2}}} \tag{77}$$

For $\frac{\nu}{\lambda^2} = 0$ we obtain well-known Euler buckling factor $\beta = 1$ (Timoshenko, 1961; Ziegler, 1977). The formula also imply that when $\nu > 0$ then the beam with $\lambda < 2\pi\sqrt{\nu}$ will not buckle. In particular for shear-lees beam with $\nu = 1$ we obtain $\lambda < 2\pi$ . This agree with result of other authors (Britvec, 1973;

Magnusson et al., 2001; Stemple, 1990). Some numerical values for $\beta$ for data provide by Humer (Humer, 2013) are given in Table 4. In Table 5 a comparisons of results for Reissne's beam and Huddelson's model of beam (Huddleston, 1972) is shown. We note that unlike the Huddelson's beam the Reisner's beam has the critical force for any $\nu < 0$.

The expression for $Y(1)=0$ is obtained by means of Eqs (19) and (43). To fulfil this boundary condition we have two possibilities. The first one is when $\sin\alpha = 0$ , which is satisfied by

$$\alpha = 0 \tag{78}$$

The second possibility is

$$-\frac{(1-\nu)\omega^2}{2\lambda^2}s + \frac{\tilde{\omega}^2}{\omega^2}\left[2\frac{E(\tilde{k})}{K(\tilde{k})} - 1\right] = 0 \tag{79}$$

In this case the beam forms a loop, that is, we have $X(1)=0$, while $\alpha$ remains indeterminate. The pair $(\omega_*, k_*)$ which determine a beam loop is obtained by solution of system of Eqs (76) and (79) . We will call $\beta = \omega_*^2/\pi^2$ the second critical factor. Figure 8 shows both critical factors as functions of $\lambda/\pi$. Some post buckling shapes of the beam are shown in Fig 9.

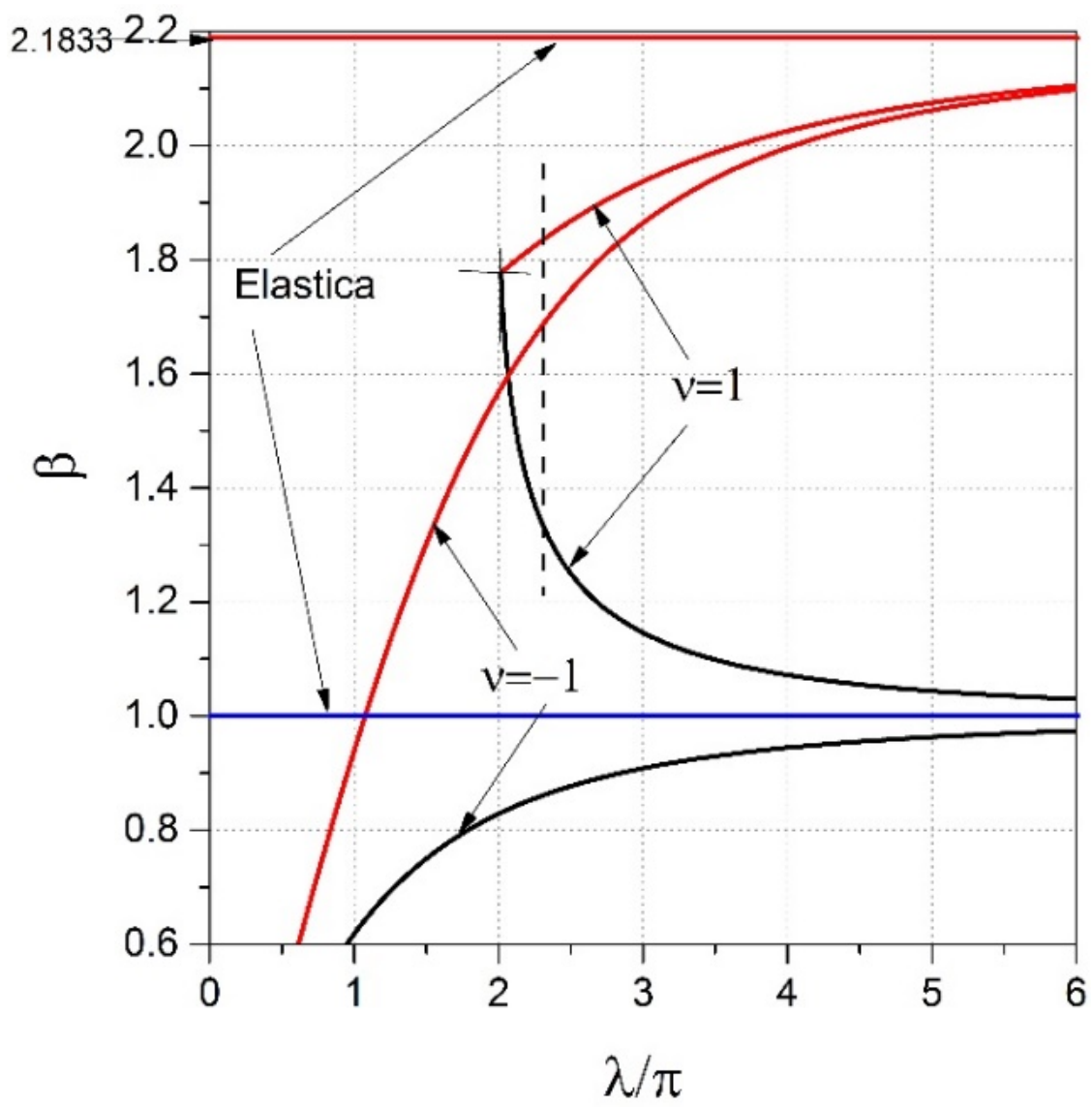

**Figure 8.** Critical factor (black line) and second critical factor (red line) versus slenderness $\lambda/\pi$ for simply supported beam. The dashed vertical line is for critical slenderness.

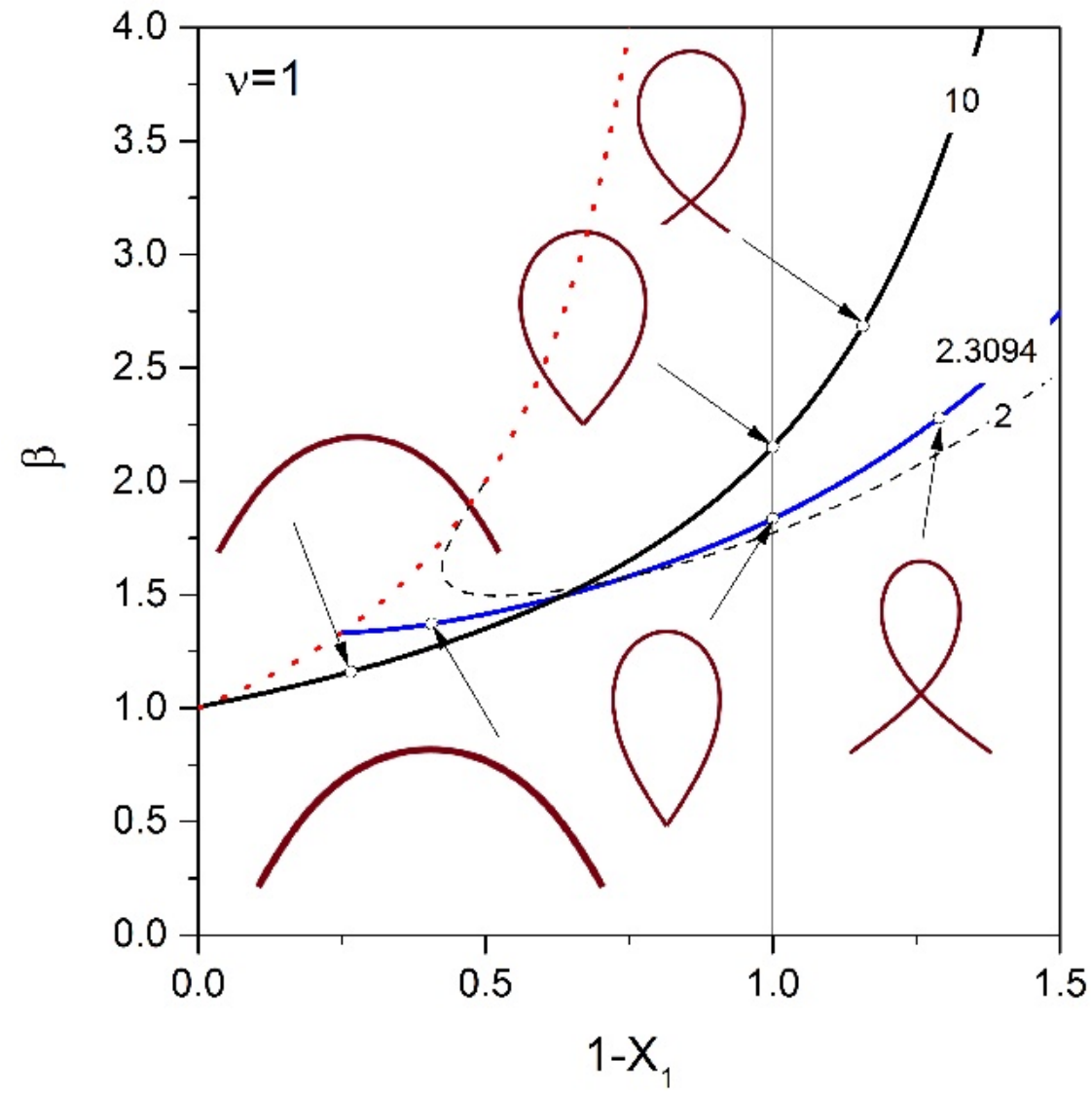

**Figure 9.** Force factor $\beta$ versus displacement of the baem movable end $1-X_1$ for simply supported beam for $\lambda/\pi \in \{2.3094, 10\}$. The deformed shape are shown for $k \in \{0.5, k_*, 0.95\}$

**Table 4.** Comparison of critical factor $\beta$ for various beam supports obtained by Humer (Humer, 2013) and present method.

| $\lambda'/\pi$ | $\nu'$ | $\lambda/\pi$ | $\nu$ | cantilever | SS | CC | CC asym | CS | Method |
|---|---|---|---|---|---|---|---|---|---|
| 5 | 1/2 | 2.8868 | $-1/3$ | 0.2476 | 0.9629 | 3.5078 | 6.1522 | 1.8728 | Humer |
| | | | | 0.247549 | 0.962912 | 3.507811 | 6.152234 | 1.872783 | Present |
| | 2 | 4.0825 | $1/3$ | 0.2513 | 1.0208 | 4.3845 | 9.8693 | 2.1268 | Humer |
| | | | | 0.251263 | 1.020842 | 4.384472 | 9.869333 | 2.126805 | Present |
| | 100 | 4.9752 | 0.9802 | 0.2525 | 1.0431 | 4.9835 | | 2.2455 | Humer |
| | | | | 0.252525 | 1.043086 | 4.983462 | | 2.245162 | Present |

**Table 5.** Comparison of critical factor $\beta$. *Diff* is relative difference with respect to present solution

| Huddleston (Huddleston, 1972) | | | present | | | |
|---|---|---|---|---|---|---|
| $EI/EA\ell^2$ | $\pi^2 EI/GA_s\ell^2$ | $\beta$ | $\lambda$ | $\nu$ | $\beta$ | Diff % |
| 0.02 | 0.25 | 0.9966 | 4.6968 | -0.1176 | 0.9523 | 4.6519 |
| 0.02 | 0.5 | 0.8438 | 3.7612 | -0.4339 | 0.8043 | 4.9111 |
| 0.02 | 0.75 | 0.7510 | 3.2276 | -0.5833 | 0.7164 | 4.8297 |
| 0.02 | 1 | 0.6860 | 2.8710 | -0.6703 | 0.6553 | 4.6849 |
| 0.03 | 0.5 | 0.9428 | 3.5210 | -0.2561 | 0.8520 | 10.6573 |
| 0.04197 | 0.5 | 1.4142 | 3.2857 | -0.0938 | 0.9264 | 52.6571 |
| 0.5 | 0.045 | * | 3.2332 | -0.0592 | 0.94962 | |

* no real solutions

### 6.2 Clamped beam

The boundary conditions for this case are

$$\phi(0)=\phi(1)=0\,,\qquad Y(1)=0 \tag{80}$$

Using Eqs (22) and (41) we find from the first two boundary conditions that $\psi(0)=\psi(1)=\alpha$. This yield equation

$$\mathrm{sn}\left(C,\tilde{k}\right)=\mathrm{sn}\left(\tilde{\omega}+C,\tilde{k}\right) \tag{81}$$

One way to satisfy this equation is to choose $\tilde{\omega}$ to be the multiple of period of *sn* and *cn* functions

$$\tilde{\omega}=4nK\left(\tilde{k}\right)\qquad \left(n=1,2,\ldots\right) \tag{82}$$

This is symmetric solution. An examples of bifurcation diagrams of Eq (82) are shown in Fig 10. Like in the case of simply supported beam we can for $\nu>0$ calculate critical slenderness. The result of calculation is $\lambda_c=8\pi\sqrt{\nu/3}$ .

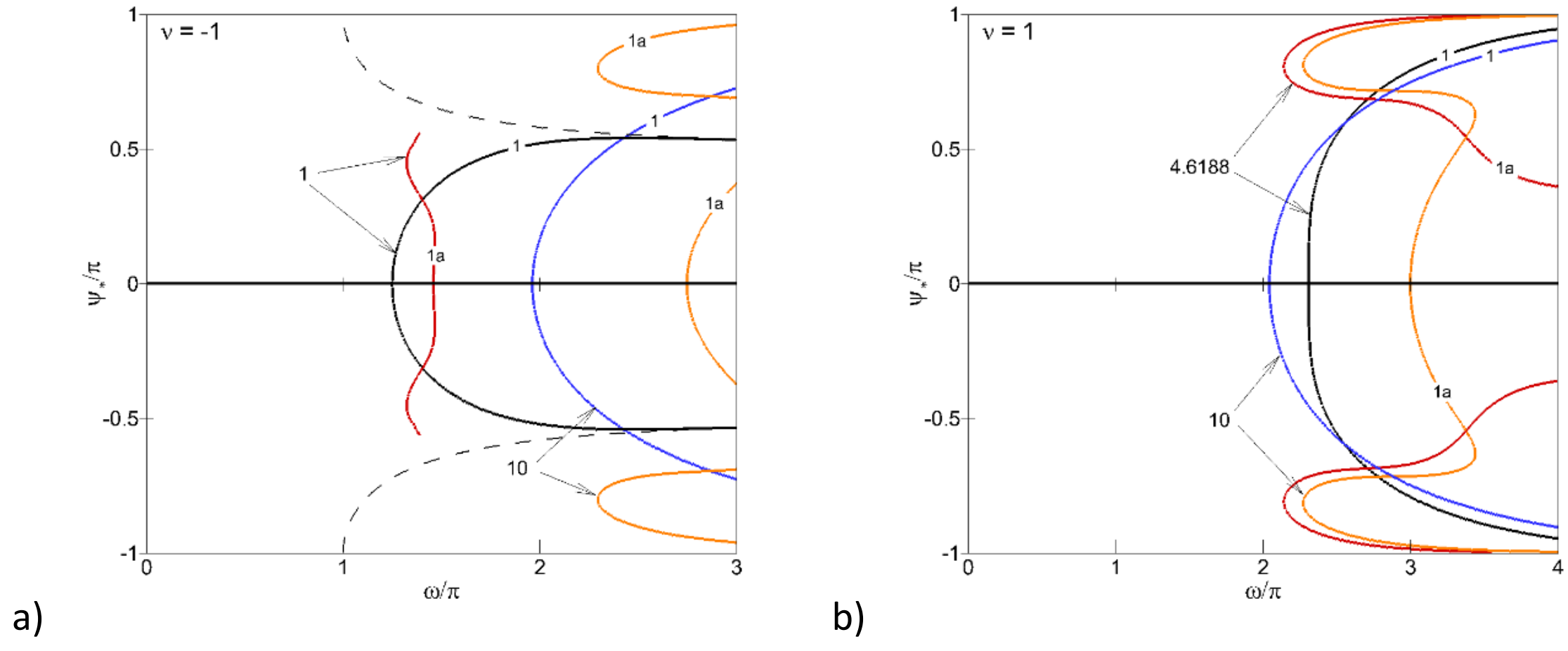

**Figure 10.** Bifurcation diagram for clamped beam for $n=1$. Label '1' is for symmetric solution, label '1a' is for asymmetric solution. a) $\nu=-1$, $\lambda/\pi\in\{1,10\}$, the dash line is boundary imposed by inequality (47) . b) $\nu=1$ , $\lambda/\pi\in\{4.6188,10\}$

Substituting $k=0$ into Eq (82) yields equation $\omega\sqrt{1+\dfrac{\nu\omega^2}{\lambda^2}}=2n\pi$ . From the solution of this equation for $n=1$ we obtain the buckling factor

$$\beta=\frac{\omega^2}{\pi^2}=\frac{8}{1+\sqrt{1-\dfrac{16\pi^2\nu}{\lambda^2}}} \tag{83}$$

For $\dfrac{\nu}{\lambda^2}=0$ this reduce to well-known Euler critical factor $\beta=4$ (Timoshenko, 1961; Ziegler, 1977). When $\nu>0$ then the beam with $\lambda\le 4\pi\sqrt{\nu}$ will not buckle.

As in the case of simply supported beam, the boundary condition $Y(1)=0$ is fulfilled in two cases. First one is $\sin\alpha=0$ with particular solution

$$\alpha = 0 \tag{84}$$

This, together with $\phi(0)=0$, gives $\mathrm{sn}(C,k)=0$ so

$$C = 0 \tag{85}$$

The second case yield to Eq (79) which is the case when $X(1)=0$, that is, when the beam forms a ribbon. We again use solution $(\omega_*, k_*)$ of Eqs (76) and (79) and (82) to form the second critical factor $\beta = \omega_*^2/\pi^2$. Both critical factors are shown in Fig 11. Some post buckling shapes of the clamped beam are show in Fig 12 an 13.

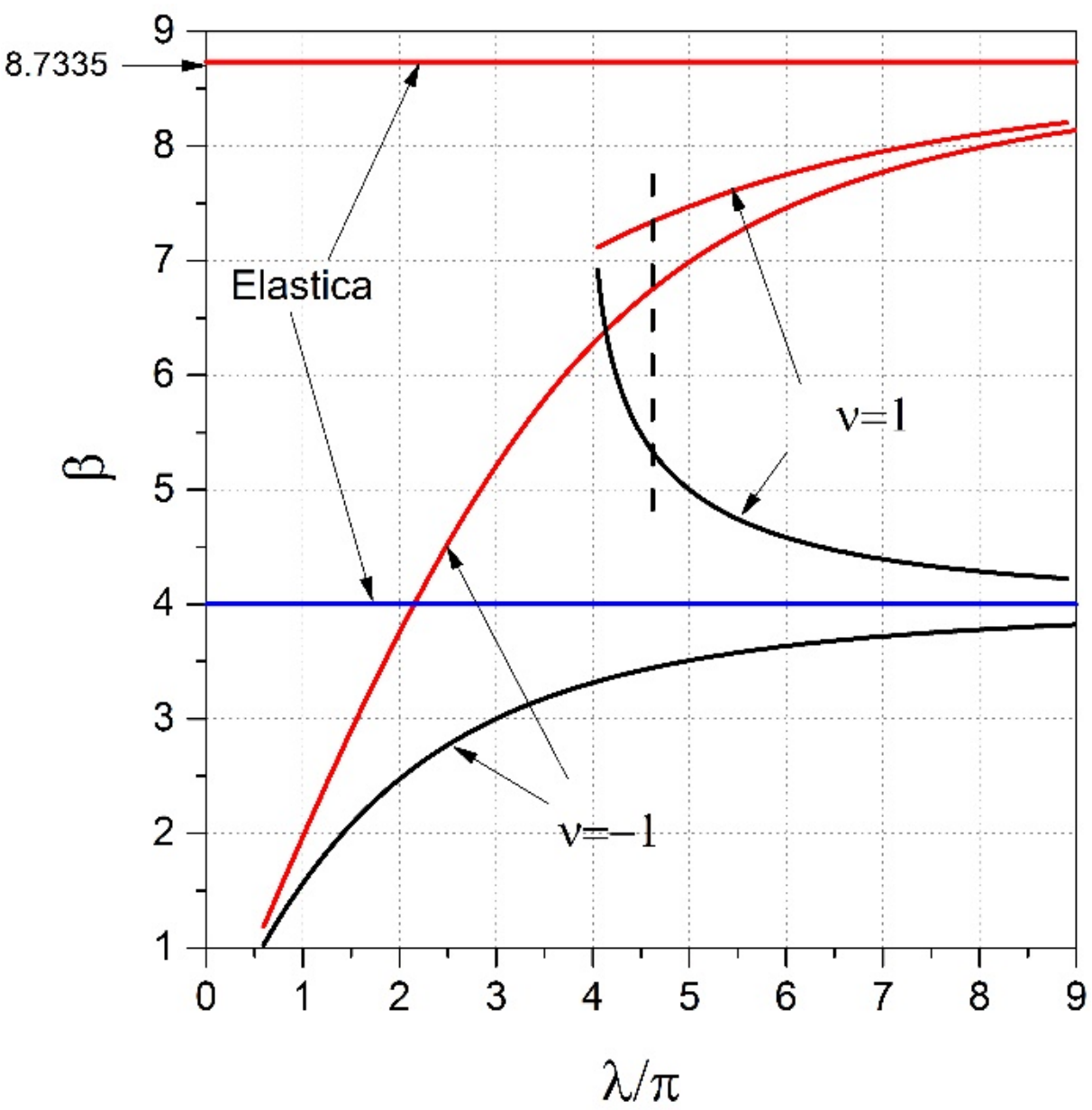

**Figure 11.** Critical factor (black line) and second critical factor (red line) versus slenderness $\lambda/\pi$ for clamped beam. The dashed vertical line is for critical slenderness.

We obtain asymmetric solution of Eq (81) by observing that $\mathrm{sn}(2K-z)=\mathrm{sn}(z)$ and $\mathrm{cn}(2K-z)=-\mathrm{cn}(z)$. Therefore $\mathrm{sn}C=\mathrm{sn}(z+C)$ yield $C+z=2K-C$ or in our case

$$C = K(\tilde{k}) - \frac{\tilde{\omega}}{2} \tag{86}$$

The equation connecting $\alpha$ and $k$ is therefore

$$\sin\frac{\alpha}{2} = k\frac{\mathrm{sn}\left(\tilde{\omega}/2 + K(\tilde{k}),\tilde{k}\right)}{\sqrt{1+m^2\mathrm{cn}^2\left(\tilde{\omega}/2+K(\tilde{k}),\tilde{k}\right)}} \tag{87}$$

If we express $\sin\alpha$ and $\cos\alpha$ by $\sin\frac{\alpha}{2}$ then the boundary condition $Y(1)=0$ become equation, not shown because of its length, connecting $k$ and $\omega$. For $k=0$ this equation reduce to

$$\omega\left(2-\frac{1-\nu}{\lambda^2}\omega^2\right)\cos\left(\frac{\omega}{2}\sqrt{1+\frac{\nu\omega^2}{\lambda^2}}\right)-4\sqrt{1+\frac{\nu\omega^2}{\lambda^2}}\sin\left(\frac{\omega}{2}\sqrt{1+\frac{\nu\omega^2}{\lambda^2}}\right)=0 \tag{88}$$

This equation can be solved for $\omega$ by standard numerical methods. For $\frac{\nu}{\lambda^2}=0$ the equation reduce to well-known characteristic equation for elastica

$$\frac{\omega}{2}\cos\frac{\omega}{2}-\sin\frac{\omega}{2}=0 \tag{89}$$

It can be easily shown that the symmetric and asymmetric solution intersect at the point where the beam form a ribbon.

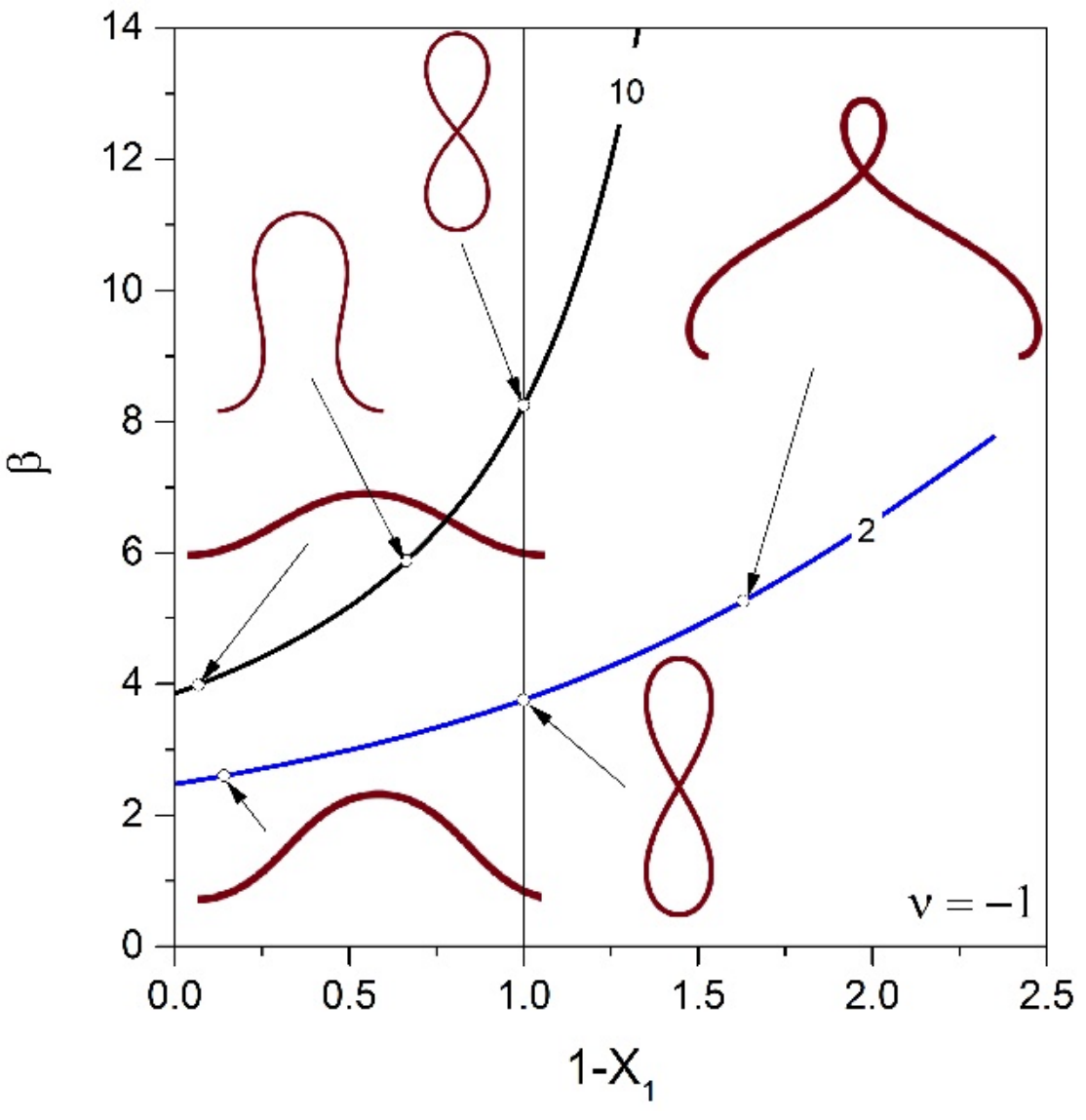

**Figure 12** Force factor $\beta$ versus displacement of movable end $1-X_1$ for clamped beam for $\lambda/\pi\in\{2,10\}$ . The deformed shape of the beam are for $k\in\{0.25,k_*,0.75\}$ . The shapes correspond to the symmetric solution.

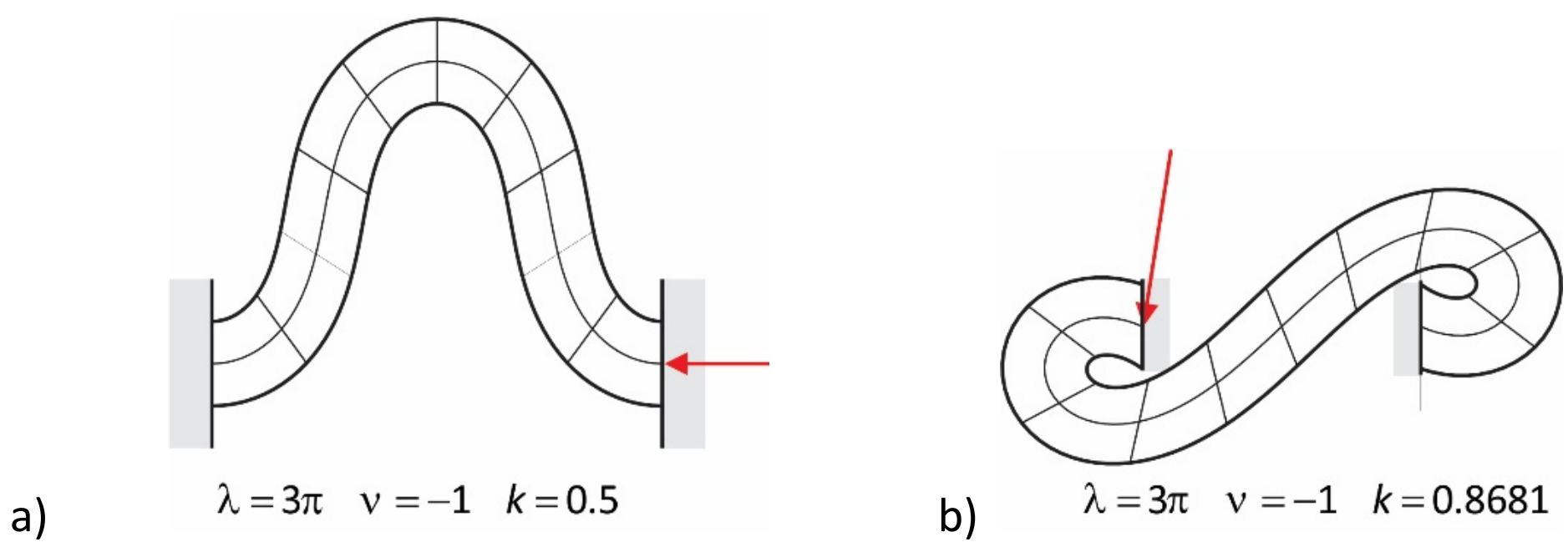

**Figure 13.** Clamped beam. Examples of symetric (a) and asymetric solution (b).

### 6.3 Case when one end is clamped and hinged

The boundary conditions for this case are

$$\phi(0)=0\,,\qquad \kappa(1)=0\,,\qquad Y(1)=0$$

By means of Eq (41) we from $\phi(0)=0$ obtain

$$\sin\frac{\alpha}{2}=k\frac{\operatorname{sn}\left(C,\tilde{k}\right)}{\sqrt{1+m^2\operatorname{cn}^2\left(C,\tilde{k}\right)}} \tag{90}$$

From the boundary condition $\kappa(1)=0$, on using Eq (42), we obtain equation $\operatorname{cn}\left(\tilde{\omega}+C,\tilde{k}\right)=0$ which is satisfied if

$$C=-\tilde{\omega}+K\left(\tilde{k}\right) \tag{91}$$

By using well known trigonometric identities we can express $\sin\alpha$ and $\cos\alpha$ through $\sin\frac{\alpha}{2}$. On using Eq (90), we then obtained from the boundary condition $Y(1)=0$ the characteristic equation which relate $k$ and $\omega$. The bifurcation diagram of this equation, not stated because it is too messy, is shown in Fig 14. From Fig 14b can be seen that also in this case we have critical slenderness. Approximate numerical value is $\lambda_c\approx 3.4\pi$ .

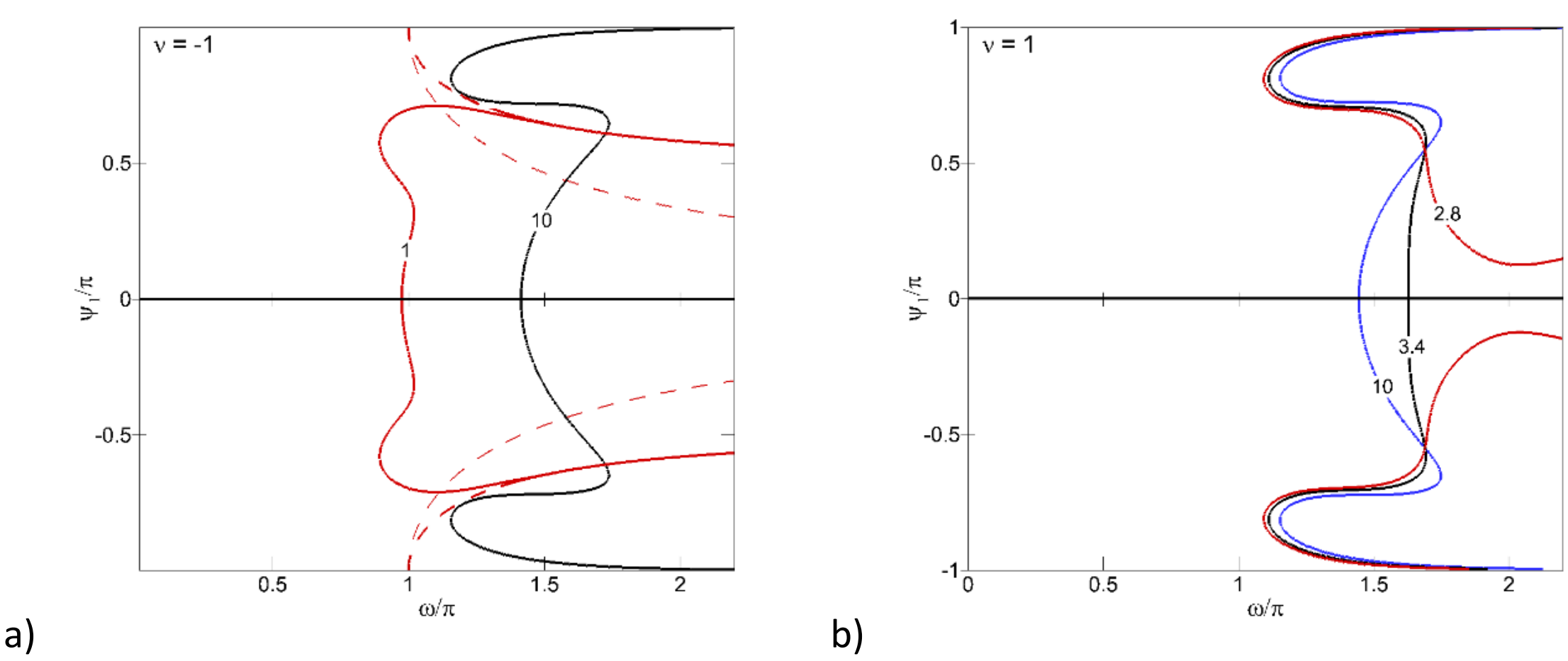

**Figure 14.** Bifurcation diagram for clamped beam. a) $\nu=-1$, $\lambda/\pi\in\{1,10\}$, the outer dash line is boundary imposed by inequality (47) while the inner dash line is boundary between real and pure complex $\tilde{k}$ . b) $\nu=1$ , $\lambda/\pi\in\{2.8,3.4,10\}$

For $k=0$ the no stated characteristic equation reduce to

$$\omega\left(2-\frac{1-\nu}{\lambda^2}\omega^2\right)\sqrt{1+\frac{\nu\omega^2}{\lambda^2}}\cos\left(\omega\sqrt{1+\frac{\nu\omega^2}{\lambda^2}}\right)-2\left(1+\frac{\nu\omega^2}{\lambda^2}\right)\sin\left(\omega\sqrt{1+\frac{\nu\omega^2}{\lambda^2}}\right)=0 \tag{92}$$

For $\frac{\nu}{\lambda^2}=0$ above equation become the well-known characteristic equation for elastica

$$\omega\cos\omega-\sin\omega=0 \tag{93}$$

Numerical solution of Eq (92) is shown in Fig 15. In the case of clamped-hinged beam we take the second critical factor to correspond to the case where horizontal component of horizontal force reach its maximal value (Fig 17). Both critical factors are shown on Fig 16. Some post buckling shapes of the beam are shown on Figs 17 and 18.

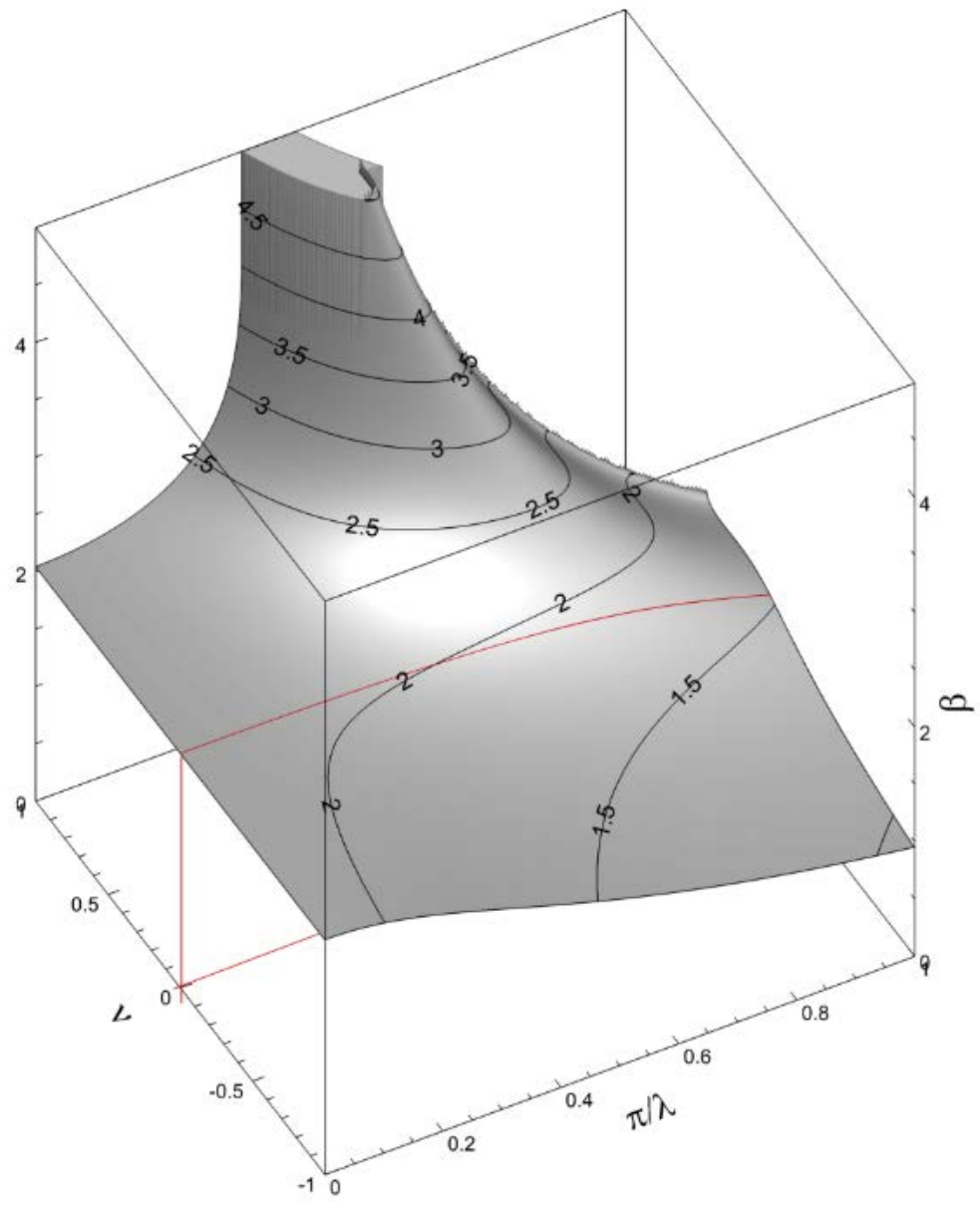

**Figure 15.** Critical force factor $\beta$ versus inverse slenderness $\pi/\lambda$ and stiffness ratio $\nu$ for clamped-hinged beam.

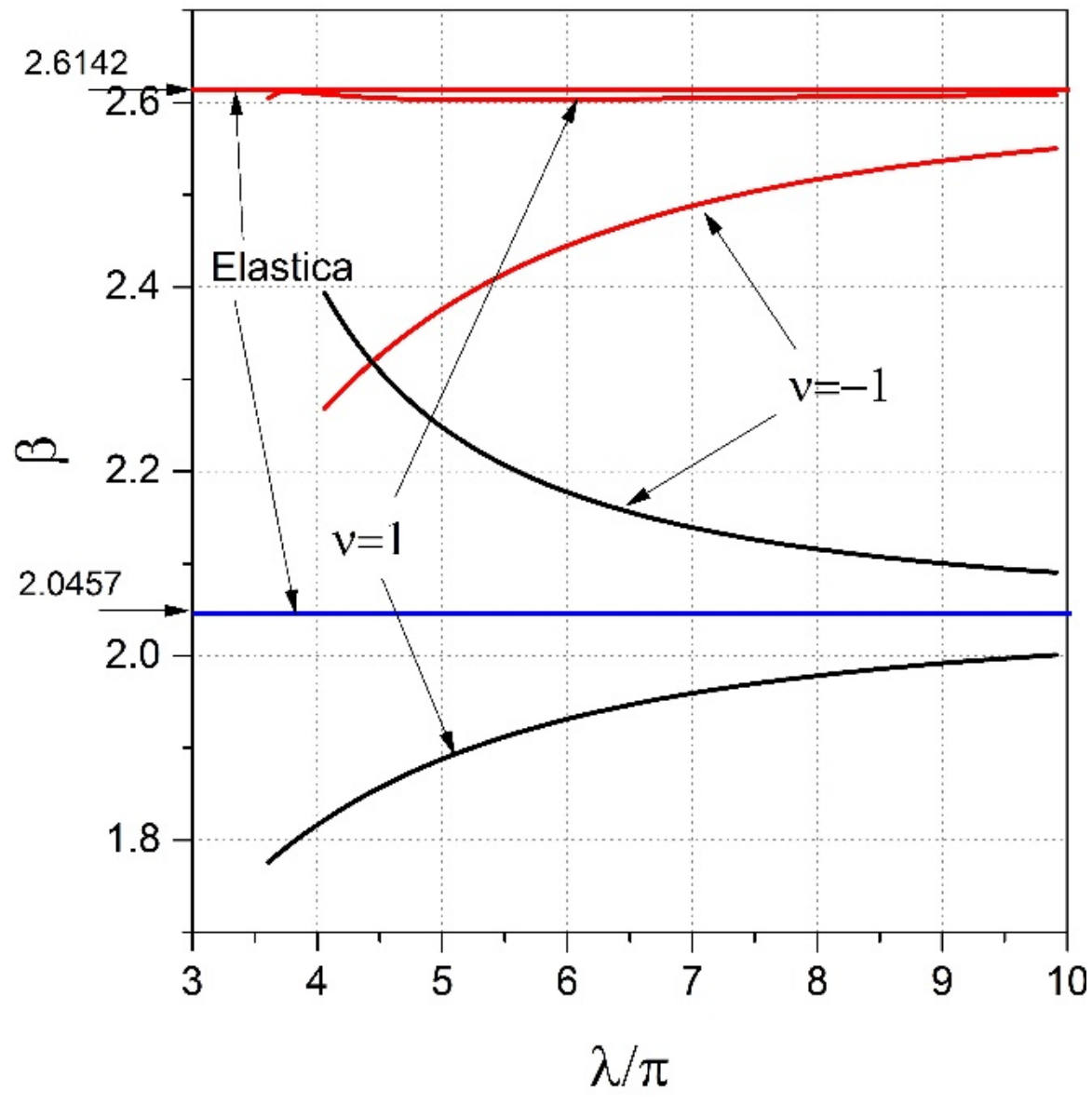

**Figure 16.** Critical factor (black line) and second critical factor (red line) versus slenderness $\lambda/\pi$ for clamped-hinged beam. Critical slenderness is about $3.4\pi$ for $\nu=1$ .

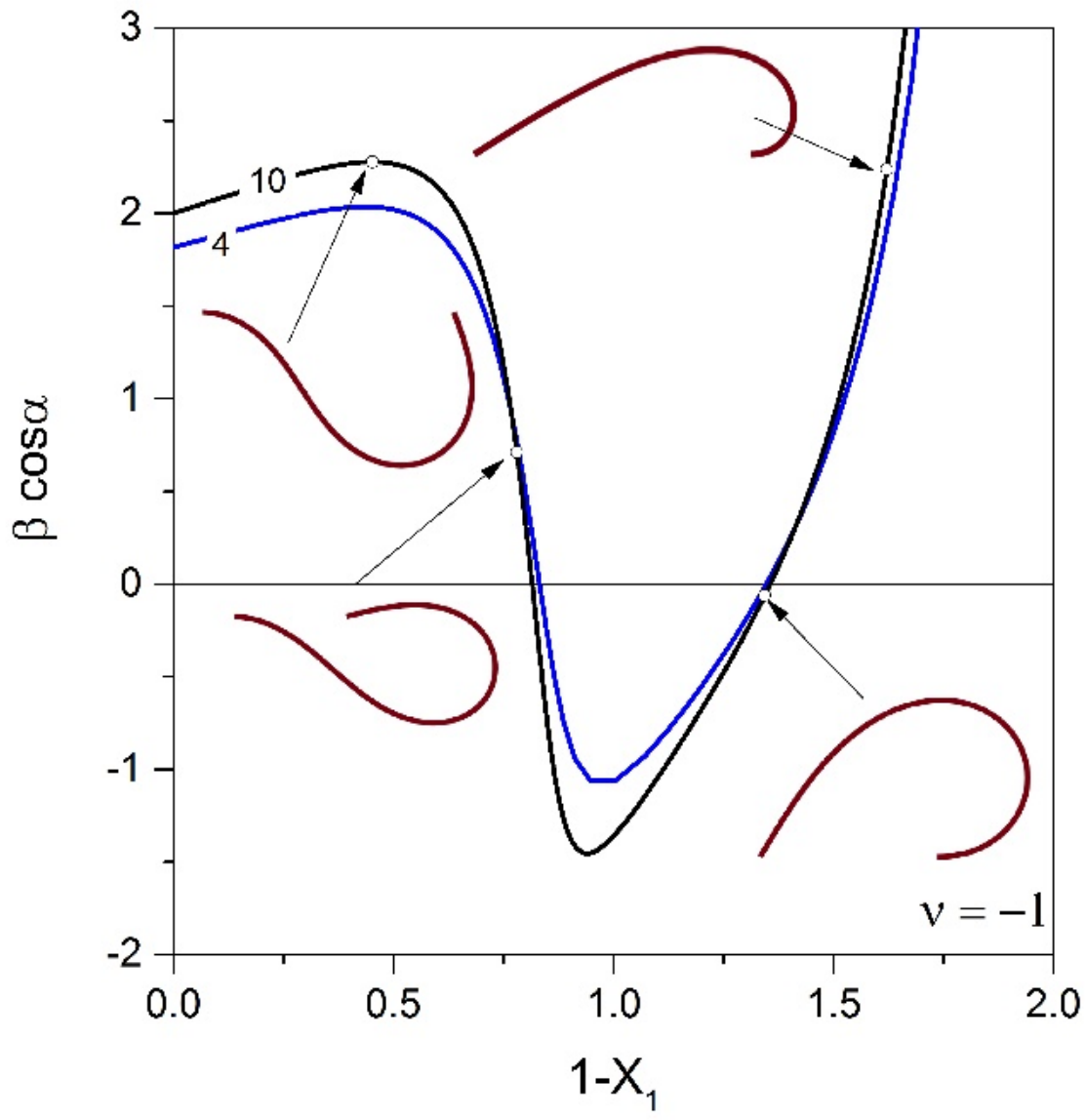

**Figure 17.** Horizontal force factor $\beta\cos\alpha$ versus displacement of movable end $1-X_1$ for clamped beam for $\lambda/\pi\in\{4,10\}$ and $\nu=-1$ . The deformed shape of the beam are for $\lambda/\pi=10$ and $\psi_1/\pi\in\{0.4567,0.6460,0.8117,0.98\}$ .

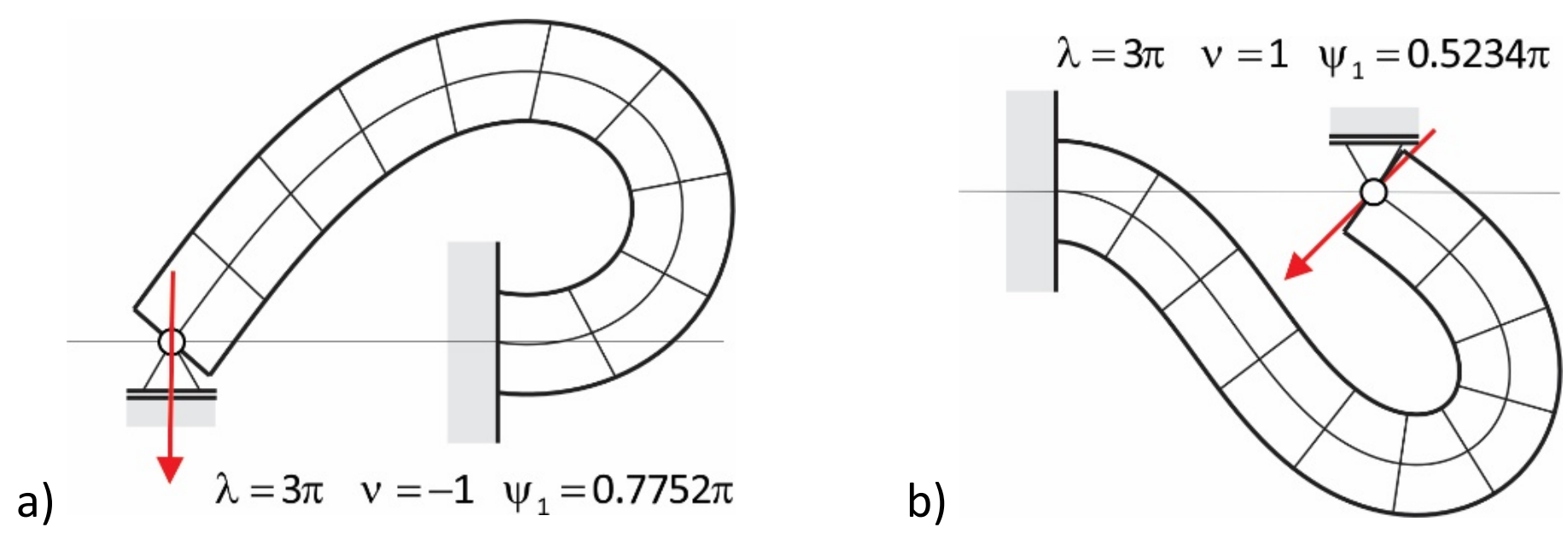

**Figure 18.** Equlibrum positions of clamped-hinged beam. Values of $\psi_1$ corespond to extrem points on bifurcation diagram.

### 6.4 Cantilever

The boundary conditions at clamped and free end are

$$\phi(0)=0\,,\qquad \kappa(1)=0 \tag{94}$$

These conditions yields two equations

$$\mathrm{sn}\left(C,\tilde{k}\right)=0\,,\qquad \mathrm{cn}\left(\tilde{\omega}+C,\tilde{k}\right)=0 \tag{95}$$

Solution of the first is

$$C=0 \tag{96}$$

and of the second

$$\tilde{\omega}=(2n-1)K(\tilde{k}) \qquad (n=1,2,\ldots) \tag{97}$$

For $k=0$ this become $2\omega\sqrt{1-\frac{\nu\omega^2}{\lambda^2}}=(2n-1)\pi$. For $n=1$ we from this equation get buckling factor

$$\beta=\frac{\omega^2}{\pi^2}=\frac{1}{2\left(1+\sqrt{1-\frac{\pi^2\nu}{\lambda^2}}\right)} \tag{98}$$

For $\frac{\nu}{\lambda^2}=0$ we have $\beta=1/4$ which is well-known Euler critical factor (Timoshenko, 1961; Ziegler, 1977). When $\nu>0$ then beams with $\lambda<\pi\sqrt{\nu}$ cannot buckle.

### 6.5 Cantilever with guided end

We assume, that the beam guide is frictionless so $\alpha=0$. The boundary conditions at clamped end and guided ends are

$$\phi(0)=0, \qquad \phi(1)=0 \tag{99}$$

From $\phi(0)=0$ we get $\mathrm{sn}(C,\tilde{k})=0$ which is satisfied by

$$C=0 \tag{100}$$

The condition $\phi(1)=0$ yield equation $\mathrm{sn}(\tilde{\omega}+C,\tilde{k})=0$ which has solution

$$\tilde{\omega}=2nK(\tilde{k}) \quad (n=1,2,\ldots) \tag{101}$$

For $k=0$ we obtain $\omega\sqrt{1-\frac{\nu\omega^2}{\lambda^2}}=n\pi$ and from this, for $n=1$, the buckling factor

$$\beta=\frac{\omega^2}{\pi^2}=\frac{2}{1+\sqrt{1-\frac{4\pi^2\nu}{\lambda^2}}} \tag{102}$$

For $\nu=0$ this reduce for well-known Euler critical factor $\beta=1$ (Timoshenko, 1961; Ziegler, 1977). When $\nu>0$ then the beam with $\lambda<2\pi\sqrt{\nu}$ will not buckle.

## 7 Conclusions

We give a new closed form solution for weightless Raissner's beam subject to end load in terms of Jacobi elliptical functions. We demonstrate that solution is efficient for numerical calculation and also for analytical investigations. In particular we demonstrate, that with present solution the derivation of the formulas for force critical factor for beam under compression force is for all discussed boundary condition almost trivial. We also derive the critical and lower limit slenderness under which the rod can't buckle for all the cases of doubly supported beam. These slenderness occurs only when $\nu > 0$.

## Appendix. The solution for moduli outside the interval [0,1)

*The case when* $k = 1$. In this case we have (Armitage and Eberlein, 2006; Carlson, 2010; Reinhardt and Walker, 2010)

$$K(1) = \infty, \quad E(1) = 1 \tag{103}$$

$$\mathrm{sn}(z,1) = Z(z,1) = \tanh z, \quad \mathrm{cn}(z,1) = \mathrm{dn}(z,1) = \mathrm{sech}\, z = \frac{1}{\cosh z}, \tag{104}$$

Introducing these relation into Eqs (41), (42) and (43) we after rearrangement obtain

$$\psi = 2\arcsin\left[\frac{\tanh(\tilde{\omega}s + C)}{\sqrt{1 + m^2 \mathrm{sech}^2(\tilde{\omega}s + C)}}\right] \tag{105}$$

$$\kappa = \frac{2\tilde{\omega}\sqrt{1+m^2}\,\mathrm{sech}(\tilde{\omega}s + C)}{1 + m^2 \mathrm{sech}^2(\tilde{\omega}s + Ck)} \tag{106}$$

$$\begin{aligned} x &= -\left[\frac{(1+\eta)\omega^2}{2\lambda^2} + 1 + m^2\right]s + \frac{2(1+m^2)}{\tilde{\omega}}\left[\frac{\tanh(\tilde{\omega}s + C)}{1 + m^2 \mathrm{sech}^2(\tilde{\omega}s + C)} - \frac{\tanh(C)}{1 + m^2 \mathrm{sech}^2(C)}\right] \\ y &= \frac{2\tilde{\omega}\sqrt{+m^2}}{\omega^2}\left[\frac{\mathrm{sech}(C)}{1 + m^2 \mathrm{sech}^2(C)} - \frac{\mathrm{sech}(\tilde{\omega}s + C)}{1 + m^2 \mathrm{sech}^2(\tilde{\omega}s + C)}\right] \end{aligned} \tag{107}$$

*The case when* $k > 1$. The real parts of elliptic integrals are given by (Carlson, 2010)(Section 19.7.3)

$$\mathrm{K}(k) = \frac{1}{k}\mathrm{K}(1/k), \quad E(k) = kE(1/k) + \frac{1}{k}(1 - k^2)\mathrm{K}(1/k) \tag{108}$$

For Jacobian elliptic functions we have the following formulas (Reinhardt and Walker, 2010) (section 22.17)

$$\mathrm{sn}(z,k) = \frac{1}{k}\mathrm{sn}(kz, 1/k), \quad \mathrm{cn}(z,k) = \mathrm{dn}(kz, 1/k), \quad \mathrm{dn}(z,k) = \mathrm{cn}(kz, 1/k) \tag{109}$$

From these formulas and the definition of *Z* function (Reinhardt and Walker, 2010) we can easily deduce the following relation

$$\mathrm{Z}(z,k) = k\mathrm{Z}(kz, 1/k) \tag{110}$$

Substituting these into Eq (41), (42) and (43) we obtain

$$\psi = 2\arcsin\left[\frac{k}{\tilde{k}}\frac{\mathrm{sn}(\tilde{k}\tilde{\omega}s + C, \tilde{k}^{-1})}{\sqrt{1 + m^2 \mathrm{dn}^2(\tilde{k}\tilde{\omega}s + C, \tilde{k}^{-1})}}\right] \tag{111}$$

$$\kappa = \pm\frac{2\tilde{\omega}k\sqrt{1+m^2}\,\mathrm{dn}\left(\tilde{k}\tilde{\omega}s+C,\tilde{k}^{-1}\right)}{1+m^2\mathrm{dn}^2\left(\tilde{k}\tilde{\omega}s+C,\tilde{k}^{-1}\right)} \tag{112}$$

$$\begin{aligned} x = &-\frac{(1-\nu)\omega^2}{2\lambda^2}s \\ &+\frac{2\tilde{\omega}}{\omega^2\tilde{k}}\left\{\left[\tilde{k}^2\left(\frac{E\left(\tilde{k}^{-1}\right)}{K\left(\tilde{k}^{-1}\right)}-\frac{1}{2}\right)+1\right]\tilde{k}\tilde{\omega}s+\tilde{k}^2\left[Z\left(\tilde{k}\tilde{\omega}s+C,\tilde{k}^{-1}\right)-Z\left(C,\tilde{k}^{-1}\right)\right]-m^2\right. \\ &\left.\times\left[\frac{\mathrm{sn}\left(\tilde{k}\tilde{\omega}s+C,\tilde{k}^{-1}\right)\mathrm{cn}\left(\tilde{k}\tilde{\omega}s+C,\tilde{k}^{-1}\right)\mathrm{dn}\left(\tilde{k}\tilde{\omega}s+C,\tilde{k}^{-1}\right)}{1+m^2\mathrm{dn}^2\left(\tilde{k}\tilde{\omega}s+C,\tilde{k}^{-1}\right)}-\frac{\mathrm{sn}\left(C,\tilde{k}^{-1}\right)\mathrm{cn}\left(C,\tilde{k}^{-1}\right)\mathrm{dn}\left(C,\tilde{k}^{-1}\right)}{1+m^2\mathrm{dn}^2\left(C,\tilde{k}^{-1}\right)}\right]\right\} \end{aligned} \tag{113}$$

$$y = \frac{2\tilde{\omega}k\sqrt{1+m^2}}{\omega^2}\left[\frac{\mathrm{dn}\left(C,\tilde{k}^{-1}\right)}{1+m^2\mathrm{dn}^2\left(C,\tilde{k}^{-1}\right)}-\frac{\mathrm{dn}\left(\tilde{k}\tilde{\omega}s+C,\tilde{k}^{-1}\right)}{1+m^2\mathrm{dn}^2\left(\tilde{k}\tilde{\omega}s+C,\tilde{k}^{-1}\right)}\right]$$

For numerical calculations the formula (111) for calculation of $\psi$ must be used by some care. Namely, it follows from Eq (112) that $\kappa$ is does not change sign. Consequently $\psi$ should be by of Eq (13) monotone function. However, Eq (111) give periodic solution. Thus, $\psi$ should be computed by $\arcsin\left(f(x)\right)$ decomposed in the following way

$$\arcsin\left(f(x)\right) = \frac{\pi}{2K}x + F\left(f(x)\right) \tag{114}$$

where

$$F\left(f(x)\right) = \arcsin\left(f(t)\right) - \frac{\pi}{2K}t \quad \text{and} \quad t = x - nK, \quad t < K \tag{115}$$

*The case when k is pure imaginary number*. For elliptic integrals we have (Carlson, 2010)

$$\mathrm{K}(ik) = k_1'K(k_1), \qquad E(ik) = (1/k_1')E(k_1) \tag{116}$$

where $i^2 = -1$ and

$$k_1 = \frac{k}{\sqrt{1+k^2}}, \qquad k_1' = \frac{1}{\sqrt{1+k^2}} \tag{117}$$

For the Jacobi elliptical functions the following relations hold (Reinhardt and Walker, 2010)

$$\mathrm{sn}(z,ik) = k_1'\frac{\mathrm{sn}\left(z/k_1',k_1\right)}{\mathrm{dn}\left(z/k_1',k_1\right)}, \qquad \mathrm{cn}(z,ik) = \frac{\mathrm{cn}\left(z/k_1',k_1\right)}{\mathrm{dn}\left(z/k_1',k_1\right)}, \qquad \mathrm{dn}(z,ik) = \frac{1}{\mathrm{dn}\left(z/k_1',k_1\right)} \tag{118}$$

By using these formulas and the definition of *Z* function (Reinhardt and Walker, 2010) we can easily derive the following relation

$$Z(z,ik) = \frac{Z\left(z/k_1'+K(k_1),k_1\right)}{k_1'} \tag{119}$$

Resulting expressions for the solution given by Eq (41), (42) and (43) are in this to extensive so we omit them.

## References

Antman, S., 1972. The theory of rods, in: Truesdell, C. (Ed.), Handbuch der Physik. Springer-Verlag, Berlin, pp. 641-703.
Antman, S.S., 2005. Nonlinear problems of elasticity, 2nd ed. Springer, New York.
Armitage, J.V., Eberlein, W.F., 2006. Elliptic functions. Cambridge University Press, Cambridge.
Batista, M., 2013. Large deflections of shear-deformable cantilever beam subject to a tip follower force. Int J Mech Sci 75, 388-395.
Batista, M., 2014. Analytical treatment of equilibrium configurations of cantilever under terminal loads using Jacobi elliptical functions. Int J Solids Struct 51, 2308-2326.
Batista, M., 2015a. On Stability of Elastic Rod Planar Equilibrium Configurations. Internation journal of solid and structures Article in press.
Batista, M., 2015b. A simplified method to investigate the stability of cantilever rod equilibrium forms. Mech Res Commun 67, 13-17.
Batista, M., Kosel, F., 2005. Cantilever beam equilibrium configurations. Int J Solids Struct 42, 4663-4672.
Britvec, S.J., 1973. The stability of elastic systems. Pergamon Press, New York,.
Carlson, B.C., 2010. Elliptic Integrals, in: Olver, F.W.J., National Institute of Standards and Technology (U.S.) (Eds.). Cambridge University Press : NIST, Cambridge ; New York, pp. xv, 951 p.
Gorski, W., 1976. A review of literature and a bibliografy on finite elastic deflections of bars. Civil Engineering Transaction, 74-85.
Goss, V.G.A., 2003. Snap buckling, writhing and loop formation in twisted rods, Center for Nonlinear Dynamics. University Collage London.
Goto, Y., Yoshimitsu, T., Obata, M., 1990. Elliptic Integral Solutions of Plane Elastica with Axial and Shear Deformations. Int J Solids Struct 26, 375-390.
Groebner, W., Hofreiter, N., 1961. Integraltafel, 3., verb., Aufl. ed. Springer, Wien,.
Hairer, E., Nørsett, S.P., Wanner, G., 1987. Solving ordinary differential equations. Springer, Berlin.
Huddleston, J.V., 1972. Effect of shear deformation on the elastica with axial strain. Internationla journal of numerical methods in engineering 4, 433-444.
Humer, A., 2013. Exact solutions for the buckling and postbuckling of shear-deformable beams. Acta Mech 224, 1493-1525.
Irschik, H., Gerstmayr, J., 2009. A continuum mechanics based derivation of Reissner's large-displacement finite-strain beam theory: the case of plane deformations of originally straight Bernoulli-Euler beams. Acta Mech 206, 1-21.
Lawden, D.F., 1989. Elliptic functions and applications. Springer-Verlag, New York.
Levyakov, S.V., 2001. States of equilibrium and secondary loss of stability of a straight rod loaded by an axial force. Journal of Applied Mechanics and Tehnical Physics 42, 321-327.
Love, A.E.H., 1944. A treatise on the mathematical theory of elasticity, 4th ed. Dover Publications, New York,.
Magnusson, A., Ristinmaa, M., Ljung, C., 2001. Behaviour of the extensible elastica solution. Int J Solids Struct 38, 8441-8457.
Pflüger, A., 1950. Stabilitätsprobleme der Elastostatik. Springer, Berlin,.
Reinhardt, W.P., Walker, P.L., 2010. Jacobian Elliptic Functions, in: Olver, F.W.J. (Ed.), NIST handbook of mathematical functions. Cambridge University Press : NIST, Cambridge ; New York, pp. xv, 951 p.
Reissner, E., 1972. One-Dimensional Finite-Strain Beam Theory - Plane Problem. Z Angew Math Phys 23, 795-804.

Saje, M., 1991. Finite-Element Formulation of Finite Planar Deformation of Curved Elastic Beams. Comput Struct 39, 327-337.
Shvartsman, B.S., 2007. Large deflections of a cantilever beam subjected to a follower force. J Sound Vib 304, 969-973.
Stemple, T., 1990. Extensional Beam-Columns - an Exact Theory. Int J Nonlinear Mech 25, 615-623.
Stoker, J.J., 1968. Nonlinear elasticity. Gordon and Breach, Science Publishers, Inc., New York.
Timoshenko, S., 1961. Theory of elastic stability, 2d ed. McGraw-Hill, New York,.
Ziegler, H., 1977. Principles of structural stability, ed.2 ed. Birkhäuser, Basel Stuttgart.